\documentclass[10pt,twocolumn,twoside]{IEEEtran}

\usepackage{ifpdf, flushend,subfigure}
\usepackage{graphicx}
\usepackage{epstopdf}
\graphicspath{{../}}
\usepackage{multirow}
\usepackage{float}
\usepackage{amsthm}
\usepackage[cmex10]{amsmath}
\usepackage{amssymb,bm}
\usepackage{mathtools}
\usepackage{graphicx}
\usepackage{ragged2e}
\usepackage{array}
\usepackage{mdwmath}
\usepackage{mdwtab}
\usepackage{eqparbox}
\usepackage{nomencl}
\usepackage{cite}
\usepackage{algorithm}
\usepackage{algpseudocode}
\usepackage{blindtext}
\usepackage{hyperref}
\usepackage{caption}
\usepackage{xcolor}
\def\bz{\mathbf{z}}

\newcounter{algsubstate}
\renewcommand{\thealgsubstate}{\alph{algsubstate}}

\algnewcommand{\Initialize}[1]{%
  \State \textbf{Initialize:}
  \Statex \hspace*{\algorithmicindent}\parbox[t]{.8\linewidth}{\raggedright #1}
}

\def\revtwo#1{{\color{black}#1}}

\usepackage{amsmath,amssymb,amsfonts}
\usepackage{graphicx}
\usepackage{textcomp}

\usepackage{amsmath,mathtools, amssymb, bbm, xspace}
\usepackage{booktabs,ragged2e}
\usepackage{multirow}
\usepackage[flushleft]{threeparttable}

\usepackage{graphicx}

\renewcommand{\vec}[1]{\ensuremath{\mathbf{#1}}}  

\def\det{\mathop{\hbox{\rm det}}}
\def\diag{\mathop{\hbox{\rm diag}}}

\def\spose#1{\hbox to 0pt{#1\hss}}

\def\text #1{\hbox{\quad#1\quad}}

\def\bbG{\mathbb{G}}

\def\nthinsp{\mskip -2   mu}

\def\K{_{\scriptscriptstyle K}}

\def\superstar{^{\raise 0.5pt\hbox{$\nthinsp *$}}}
\def\SUPERSTAR{^{\raise 0.5pt\hbox{$*$}}}

\def\lamstarT {\lambda^{\raise 0.5pt\hbox{$\nthinsp *$}T}}

\def\hbar{\skew{4.2}\bar h}

		\def\bkE{{\rm I\kern-.17em E}}
		\def\bk1{{\rm 1\kern-.17em l}}
		\def\bkD{{\rm I\kern-.17em D}}
		\def\bkR{{\rm I\kern-.17em R}}
		\def\bkP{{\rm I\kern-.17em P}}
		\def\bkY{{\bf \kern-.17em Y}}
		\def\bkZ{{\bf \kern-.17em Z}}

		\def\bc{\begin{center}}
		\def\be{\begin{enumerate}}
		\def\bi{\begin{itemize}}
		\def\ec{\end{center}}
		\def\ee{\end{enumerate}}
		\def\ei{\end{itemize}}
		\def\es{\end{small}}
		\def\eS{\end{slide}}

	\def\cp2problem#1#2#3#4{\fbox
		 {\begin{tabular*}{0.9\textwidth}
			{@{}l@{\extracolsep{\fill}}l@{\extracolsep{6pt}}l@{\extracolsep{\fill}}c@{}}
				#1 & & $#4 $ 
			\end{tabular*}}}

		\def\bkE{{\rm I\kern-.17em E}}
		\def\bk1{{\rm 1\kern-.17em l}}
		\def\bkD{{\rm I\kern-.17em D}}
		\def\bkR{{\rm I\kern-.17em R}}
		\def\bkP{{\rm I\kern-.17em P}}
		
		\def\bkZ{{\bf{Z}}}

\newcommand {\beeq}[1]{\begin{equation}\label{#1}}
\newcommand {\eeeq}{\end{equation}}
\newcommand {\bea}{\begin{eqnarray}}
\newcommand {\eea}{\end{eqnarray}}

\def\texitem#1{\par\smallskip\noindent\hangindent 25pt
               \hbox to 25pt {\hss #1 ~}\ignorespaces}

\newcommand{\beq}{\begin{equation}}
\newcommand{\eeq}{\end{equation}}
\newcommand{\beqn}{\begin{eqnarray}}
\newcommand{\eeqn}{\end{eqnarray}}
\newcommand{\beqno}{\begin{eqnarray*}}
\newcommand{\eeqno}{\end{eqnarray*}}
\newcommand{\bma}{\begin{displaymath}}
\newcommand{\ema}{\end{displaymath}}
\newcommand{\bnu}{\begin{enumerate}}
\newcommand{\enu}{\end{enumerate}}
\newcommand{\bce}{\begin{center}}
\newcommand{\ece}{\end{center}}
\newcommand{\btb}{\begin{tabular}}
\newcommand{\etb}{\end{tabular}}

\newcommand{\col}{\mathrm{col}}

\def\D{{\mathsf{D}}}

\def\E{{\mathcal{E}}}
\def\K{{\mathsf{K}}}

\def\bp{{\boldsymbol{p}}}
\def\bx{{\boldsymbol{x}}}

\def\by{{\boldsymbol{y}}}
\def\bz{{\boldsymbol{z}}}
\def\bu{{\boldsymbol{u}}}
\def\bv{{\boldsymbol{v}}}
\def\bL{{\mathbf{L}}}

\def\bF{{\mathbf{F}}}
\def\b1{{\mathbf{1}}}

\def\bdeta{\boldsymbol{\eta}}
\def\bdbeta{\boldsymbol{\beta}}
\def\bda{\boldsymbol{\alpha}}

\def\cNinik{{\mathcal{N}^{\rm in}_{ik}}}
\def\cNoutik{{\mathcal{N}^{\rm out}_{ik}}}
\def\cNini{{\mathcal{N}^{\rm in}_{i}}}
\def\cNouti{{\mathcal{N}^{\rm out}_{i}}}

\def\diag{{\rm diag}}

\usepackage[normalem]{ulem} 

\newcommand{\lmax}[1]{{\lambda_{\max}\{#1\}}}

\newtheorem{theorem}{Theorem}
\newtheorem{definition}{Definition}

\newtheorem{lemma}{Lemma}
\newtheorem{remark}{Remark}

\newtheorem{assumption}{Assumption}
\newtheorem{corollary}{Corollary}

\newcommand{\g}{{\mathbf{g}}}

\newcommand{\Diag}{{\mathrm{Diag}}}
\newcommand{\di}{{\mathrm{diag}}}

\newcommand{\ra}{{r_\mathrm{a}}}

\newcommand{\one}{{\mathbf{1}}}
\newcommand{\zero}{{\mathbf{0}}}

\definecolor{myBlue}{rgb}{0.80,0.85,1.00}
\definecolor{myYellow}{rgb}{0.951,1.000,0.547}

\def\la{{\langle}}
\def\ra{{\rangle}}

\def\bit{\begin{itemize}}
\def\eit{\end{itemize}}
\def\BEAS{\begin{eqnarray*}}
\def\EEAS{\end{eqnarray*}}

\def\diag{\rm diag}
\def\ones{{\bf 1}}

\def\re{{\mathbb R}}

\def\a{\alpha}
\def\b{\beta}

\def\g{\gamma}

\IEEEoverridecommandlockouts     
\begin{document}

\title{Distributed Nash Equilibrium Seeking over Time-Varying Directed Communication Networks }

\author{Duong Thuy Anh Nguyen, \IEEEmembership{Student Member, IEEE},
Duong Tung Nguyen, \IEEEmembership{Member, IEEE}, \\and Angelia Nedi\'c, \IEEEmembership{Member, IEEE}
\thanks{The authors are with the School of Electrical, Computer and Energy Engineering, Arizona State University, Tempe, AZ, United States. 
Email: \{dtnguy52,~duongnt,~Angelia.Nedich\}@asu.edu. 
This material is based in part upon work supported by DOD-NAVY Office of Naval Research award N00014-21-1-2242, the NSF award CCF-2106336, and the ARPA-H award SP4701-23-C-0074.
\textit{Corresponding Author}: Duong Thuy Anh Nguyen.} 
}
\maketitle

\begin{abstract}
This paper proposes a distributed algorithm to find the Nash equilibrium in a class of non-cooperative convex games with partial-decision information. Our method employs a distributed projected gradient play approach alongside consensus dynamics, with individual agents minimizing their local costs through gradient steps and local information exchange with neighbors via a time-varying directed communication network. Addressing time-varying directed graphs presents significant challenges. Existing methods often circumvent this by focusing on static graphs or specific types of directed graphs or by requiring the stepsizes to scale with the Perron-Frobenius eigenvectors. In contrast, we establish novel results that provide a contraction property for the mixing terms associated with time-varying row-stochastic weight matrices. Our approach explicitly expresses the contraction coefficient based on the characteristics of the weight matrices and graph connectivity structures, rather than implicitly through the second-largest singular value of the weight matrix as in prior studies. The established results facilitate proving geometric convergence of the proposed algorithm and advance convergence analysis for distributed algorithms in time-varying directed communication networks. Numerical results on a Nash-Cournot game demonstrate the efficacy of the proposed method. 
\end{abstract}


\printnomenclature
\allowdisplaybreaks

\section{Introduction} \label{intro}
Game theory presents a systematic approach to comprehending decision-making in strategic situations involving multiple agents striving to optimize their individual, yet interdependent, objectives. Non-cooperative games, in particular, have received considerable attention across various engineering domains such as wireless networks \cite{zhan12}, electricity markets \cite{BasharSG}, power systems \cite{Scutaricdma}, sensor networks \cite{GadjovTCSN2023,Stankovic_TAC_2012}, and crowdsourcing \cite{wiopt23}. The Nash equilibrium (NE) concept plays a crucial role in identifying stable and desirable solutions, 
representing a joint action from which no agent has an incentive to unilaterally deviate.
Consequently, developing efficient NE-seeking algorithms has become increasingly imperative. 

In non-cooperative games, each agent's payoff is determined by its actions and the actions of other agents. The existing body of work, utilizing best-response or gradient-based methods, requires each agent to possess complete information regarding the actions of its competitors to search for NE \cite{Belgioioso2018,Yi_Pavel_splitting_2019}. However, this assumption of full-decision information is often impractical in real-world engineering systems \cite{Salehisadaghiani2014}, as seen in the Nash-Cournot competition \cite{Bimpikis2019}. 
Recent research has thus shifted focus towards fully distributed algorithms that rely solely on local information to compute the NE (i.e.,  partial-decision information setting \cite{GadjovTCSN2023}).These algorithms are predominantly built upon projected gradient approaches and consensus dynamics, and exist in both continuous-time \cite{Ye_Hu_Consensus_TAC_2017,Gadjov2019} and discrete-time \cite{Koshal2012,Tatiana2020,Nguyen2023AccGame,Salehisadaghiani_Wei_Pavel_AUT_2019,Tatiana2018,Tatarenko2021,Bianchi2020NashES,Bianchi_2021} forms. 

Reference~\cite{Salehisadaghiani2014} presents a gradient-based gossip algorithm for distributed NE seeking, which converges almost surely to an NE under strict convexity, Lipschitz continuity, and bounded gradient assumptions, employing a diminishing stepsize. With the additional assumption of strong convexity, convergence to an $O(\a)$ neighborhood of the NE is ensured with a constant stepsize $\a$. In~\cite{Salehisadaghiani_Wei_Pavel_AUT_2019}, an inexact-ADMM algorithm 
is developed, and its convergence rate of $o(1/k)$ is established for a fixed stepsize under the co-coercivity assumption of the game mapping. 
Reference~\cite{Tatiana2018} proposes Acc-GRANE, an accelerated version of the gradient play algorithm for solving variational inequalities. The analysis is based on the strong monotonicity of an augmented mapping, which integrates both the gradients of the cost functions and the communication settings. 
Although this algorithm is restricted to a subclass of games, reference \cite{Tatarenko2021} shows that assuming restricted strong monotonicity of the augmented mapping enables Acc-GRANE to apply to a broader class of games and achieve geometric convergence to an NE. Alternatively, leveraging contractivity properties of doubly stochastic matrices, \cite{Tatiana2020} develops a distributed gradient-based scheme whose convergence properties are independent of augmented mapping. \textit{However, all these methods are designed for undirected communication networks.}

Handling time-varying directed graphs is essential in many applications where communication networks frequently change due to user mobility, communication delays, or straggler effects. Additionally, information flow may be directed by factors such as unilateral transmission capabilities, heterogeneous power levels, or bandwidth limitations.
Early works \cite{Koshal2012} and \cite{Koshal2016} consider aggregative games over time-varying, jointly connected, and undirected graphs, with this result extended in~\cite{Grammatico2021} to games with coupling constraints. In~\cite{Farzad2019}, an asynchronous gossip algorithm is developed to find an NE over a directed graph, assuming each agent can update all estimates of agents interfering with its cost function. In~\cite{Bianchi_2021}, a projected pseudo-gradient algorithm is proposed for time-varying directed graphs with weight-balanced mixing matrices. However, constructing weight-balanced matrices, even in a directed static graph, is non-trivial and computationally expensive~\cite{Gharesifard2012}.
The weight-balanced assumption is relaxed in~\cite{Bianchi2020NashES}, where a modified algorithm is proposed for a static directed graph. However, this relaxation necessitates the computation of the Perron-Frobenius (PF) eigenvector of the adjacency matrix, a requirement further relaxed in \cite{bianchi2024end}. 
Deviating from monotonicity assumptions, reference \cite{Nguyen2023RowGame} establishes geometric convergence towards an NE under a locally verifiable diagonal dominance property of the game mapping; 
additionally, explicit bounds for constant stepsize are provided independent of the communication structure and computable in a totally decentralized manner. Nevertheless, \cite{Nguyen2023RowGame} considers a static directed network, and its extension to time-varying networks is still an open research question.

\textbf{Contributions.}
This paper addresses NE seeking for non-cooperative games in a partial-decision information setting, where each agent has access only to their own cost function and action set, and engages in nonstrategic information exchange with neighboring agents. Our contributions include: \\
\textbf{\textit{1) Distributed NE seeking algorithm in time-varying directed communication networks:}} We propose a distributed algorithm for NE computation over \textit{time-varying directed communication networks}. This approach extends beyond previous works that restrict communication networks to \textit{static} undirected graphs \cite{Tatiana2018,Salehisadaghiani_Wei_Pavel_AUT_2019,Tatiana2020,Tatarenko2021}, \textit{static} directed graphs \cite{Bianchi2020NashES,bianchi2024end}, or \textit{undirected/weight-balanced directed} time-varying graphs \cite{Koshal2016,Bianchi_2021}. Additionally, unlike previous works on directed graphs, such as \cite{Bianchi2020NashES}, which require scaling stepsizes with the PF eigenvector--a process that can be impractical and computationally expensive, especially in distributed settings--our algorithm eliminates the need for such computations.\\
\textbf{\textit{2) Novel contraction property and convergence analysis:}} Our convergence analysis directly addresses the complexities that arise from the loss of properties associated with doubly-stochastic matrices when dealing with directed graphs and the dynamic nature of mixing matrices in time-varying graphs. We employ time-varying weighted averages and norms, with weights defined based on stochastic vector sequences that are linked to the sequences mixing matrices. We introduce two novel results, namely Lemma~\ref{lem-normlincomb} and Lemma~\ref{lem-xdiverse}, and establish a new contraction property for row-stochastic matrices linked to strongly connected, time-varying directed graphs, detailed in Lemma~\ref{lemma-lemma6PushPull}. Unlike previous works that characterizes the contraction coefficient in terms of the second-largest singular value of the weight matrix \cite{Tatiana2020,Bianchi2020NashES}, our approach provides an explicit coefficient based on graph connectivity. Under the commonly-used strong convexity and Lipschitz continuity assumptions \cite{Salehisadaghiani_Wei_Pavel_AUT_2019,Tatiana2018,Tatarenko2021,Bianchi2020NashES,Bianchi_2021,Nguyen2023AccGame}, we show that our algorithm achieves geometric convergence to the NE. Our analysis does not rely on the augmented mapping as in previous works \cite{Tatiana2020, Tatarenko2021}.\\
\textbf{\textit{3) Generality of proposed approach:}} Our algorithm generalizes several existing methods, including those in \cite{Tatiana2020,Bianchi2020NashES,Bianchi_2021} as special cases. 
Our novel results extend to convergence analysis in other settings involving time-varying directed networks, including these special cases; see Remark \ref{rem-generality}.\\
\textbf{\textit{4) Performance evaluation:}} Simulation results for a Nash-Cournot game with time-varying directed communication networks demonstrate the performance and effectiveness of the proposed algorithm and its advantages over prior works.

The paper is organized as follows: Section~\ref{sec:formu} outlines the problem formulation; Section~\ref{sec:algo} presents the distributed algorithm; Sections~\ref{sec:basicre} and~\ref{sec:conv} cover auxiliary results and convergence analysis, respectively; Section~\ref{sec:simulation} showcases simulation results; and Section~\ref{sec:conc} concludes with key points.

\section{Notations and Terminologies}\label{sec:nota}
All vectors are viewed as column vectors unless stated otherwise and 
the time index is denoted by $k$. For $u\in \mathbb{R}^n$, $u'$ is the transpose of $u$.  
 We write $\zero$ and $\one$ to denote the vectors with all entries equal to $0$ and $1$, respectively. 
The $i$-th entry of a vector $u$ is denoted by $u_i$, while it is denoted by $[u_k]_i$ for a time-varying vector $u_k$. We 
denote $\min(u)=\min_i u_i$ and $\max(u)=\max_i u_i$. 
We write $u>\zero$ to indicate that the vector $u$ has positive entries. A vector is said to be a stochastic vector if its entries are nonnegative and sum to $1$. 	

For a set $S$ with finitely many elements, we use $|S|$ to denote its cardinality. To denote the $ij$-th entry of a matrix $A$, we write $A_{ij}$, and we write $[A_k]_{ij}$ when the matrix is time-dependent. For any two matrices $A$ and $B$ of the same dimension, we write $A\le B$ to denote that $A_{ij}\le B_{ij}$, $\forall i, j$. 
A matrix is said to be nonnegative if all its entries are nonnegative. For a nonnegative matrix $A$, we 
denote ${\min}^{+}(A)=\min_{\{ij:A_{ij}>0\}} A_{ij}$. A nonnegative matrix is said to be row-stochastic if each row's entries sum to $1$.

The largest 
eigenvalue in modulus of a matrix $A$ is denoted as $\lmax{A}$.
For matrix $A\in\re^{n\times n}$, we use $\di(A)$ to denote its diagonal vector, i.e. $\di(A) = (a_{11},\ldots, a_{nn})'$.
For any vector $u\in\re^n$ we  use $\Diag(u)$ to denote the diagonal matrix with the vector $u$ on its diagonal. Given a vector $\pi\in\re^m$ with positive entries $\pi_1,\ldots,\pi_m$, the $\pi$-weighted inner product and $\pi$-weighted norm are defined, respectively, as follows:
\begin{center}
    $\la \bu,\bv\ra_{\pi}=\sum_{i=1}^m \pi_i\la u_i,v_i \ra \quad$and$\quad\|\bu\|_{\pi}=\sqrt{\sum_{i=1}^m \pi_i\|u_i\|^2},$
\end{center}
where $\bu=[u_1,\ldots,u_m]', \bv=[v_1,\ldots,v_m]' \in \re^{m\times n}$, and $u_i,v_i\in\re^n$. Also, let $\col\left( (u_i)_{i \in [m]}\right) = [u_1',\ldots,u_m']' \in\re^{mn}$, which denotes the column vector formed by vertically stacking the individual column vectors $u_i$ for $i \in[m]$. When $\pi = \one$, we simply write $\la \bu,\bv\ra$ and $\|\bu\|$, for which we have:
\begin{align}\label{eq-NormIneq}   \tfrac{1}{\sqrt{\max(\pi)}}\|\bu\|_{\pi}
\le \|\bu\| 
  \le \tfrac{1}{\sqrt{\min(\pi)}}\|\bu\|_{\pi}.
\end{align}
The Cauchy–Schwarz inequality gives: 
$\la \bu,\bv\ra_{\pi}  \le\|\bu\|_{\pi}\|\bv\|_{\pi}$.
 
We consider real normed space $E$, which is either the space of real vectors $E = \re^n$ or the space of real matrices $E = \re^{n\times n}$. We use $\Pi_\Omega [u]$ to denote the projection of $u \in E$ to a set $\Omega \subseteq E$. A mapping $g:E\to E$ is said to be \textit{strongly monotone} on a set $Q\subseteq E$ with the constant $\mu>0$, if $\langle g(u)-g(v), u - v \rangle\ge\mu\|u - v\|^2$ for any $u,v\in Q$.
A mapping
$g:E\to E$ is said to be \textit{Lipschitz continuous} 
on a set $Q\subseteq E$
with the constant $L>0$, if $\|g(u)-g(v)\|\le L\|u - v\|$.

We let $[m]$ denote the set $\{1,\ldots,m\}$ for an integer $m\ge 1$. A directed graph $\bbG=([m],\E)$ is specified by the set of edges $\E\subseteq [m]\times[m]$ of ordered pairs of nodes. 
Given two distinct nodes $j, \ell\in [m]$ ($j\ne \ell$), a \textit{directed path} from node $j$ to node $\ell$ 
in the graph $\bbG$ is a finite (ordered) sequence of edges $\{(i_0,i_1),\ldots,(i_{t-1},i_t)\}$ passing through distinct nodes, where $i_0=j,~i_t=\ell$, and $(i_{s-1},i_s)\in\E$ for all $s=1,\ldots,t$. 

\begin{definition}[Graph Connectivity]
A directed graph is \textit{strongly connected} if there is a directed path from any node to all the other nodes in the graph.
\end{definition}


The length of a path is the number of edges in the path.


\begin{definition} [Graph Diameter]\label{def-diam}
The diameter of a strongly connected directed graph $\bbG$ is the length of the longest path in the collection of all shortest directed paths connecting all ordered pairs of distinct nodes in $\bbG$. 
\end{definition}

We denote the diameter of the graph $\bbG$ by $\D(\bbG)$. Let $\bp_{j\ell}$ denote a shortest directed path from node $j$ to node $\ell$, where $j\ne \ell$. Special collections of such paths, referred to as the shortest-path covering of the graph, are defined as follows:


{
\begin{definition} [Shortest-Path Graph Covering]\label{def-sp-covering}
A collection $\mathcal{P}=\{\bp_{j\ell}\mid j,\ell\in[m], j\ne\ell\}$ of directed paths in $\bbG$ is called a shortest-path graph covering. 
\end{definition}}

Denote by $\mathcal{S}(\bbG)$  the collection of all possible shortest-path coverings of the graph $\bbG$. Given a shortest-path covering $\mathcal{P}\in\mathcal{S}(\bbG)$ and 
an edge $(j,\ell) \!\in\! \E$, the \textit{utility of the edge} $(j,\ell)$ with respect to the covering $\mathcal{P}$ is the number of shortest paths in $\mathcal{P}$ that pass through the edge $(j,\ell)$. 
Define $\K(\mathcal{P})$ as the maximum edge-utility in $\mathcal{P}$ taken over all edges in the graph, i.e.,
\begin{align*}
\textstyle \K(\mathcal{P})\!=\!\max_{(j,\ell)\in\E}\sum_{\bp\in\mathcal{P}}\chi_{\{(j,\ell)\in\bp\}},
\end{align*}
where $\chi_{\{(j,\ell)\in\bp\}}$ is an indicator function that equals $1$ when $(j,\ell)\in\bp$ and $0$ otherwise.

\begin{definition} [Maximal Edge-Utility] \label{def-edgeut}
For a strongly connected directed graph $\bbG\!=\!([m],\E)$, the maximal edge-utility $\K(\bbG)$ is defined as the maximum value of $\K(\mathcal{P})$ over all possible shortest-path coverings $\mathcal{P}\in\mathcal{S}(\bbG)$, i.e.,
\[\textstyle \K(\bbG)=\max_{\mathcal{P}\in\mathcal{S}(\bbG)}\K(\mathcal{P}).\]
\end{definition}


As an example, consider a directed-cycle graph $\bbG = ([m], \E)$. Then, \revtwo{$\mathsf{K}(\bbG) = \frac{m(m-1)}{2}$}.

Given a directed graph $\bbG=([m],\E)$, we define the in-neighbor and out-neighbor set for every agent $i$, as follows:
\[\cNini\!=\!\{j\!\in\![m] \mid (j,i)\!\in\!\E\} \!\!\!\text{and}\!\!\! \cNouti\!=\!\{\ell\!\in\![m] \mid (i,\ell)\!\in\!\E\}.\]

When the graph varies over time, we use a subscript to indicate the time instance. For example, $\E_k$ denotes the edge-set of a graph $\bbG_k$, while $\cNinik$ and $\cNoutik$ denote the in-neighbors and the out-neighbors of a node $i$, respectively. 

\section{Problem Formulation}
\label{sec:formu}
We consider a non-cooperative game between $m$ agents. For each agent $i\in[m]$, let $J_i(\cdot)$ be the cost function, $X_i\subseteq \mathbb{R}^{n_i}$ be the action set and $x_i\in X_i$ be the action of the agent. 
Define the joint action vector (action profile) as $x \coloneqq  \col( (x_i)_{i \in [m]})$, which aggregates the decisions of all agents. This vector belongs to the joint decision set $X \coloneqq X_1\times \ldots \times X_m \subseteq \mathbb{R}^{n}$, where $\textstyle n\coloneqq \sum_{i=1}^m n_i$ is the size of $x$.
Each function $J_i(x_i,x_{-i})$ depends on the agent's own action $x_i$ and the joint action $x_{-i}$ of all agents except agent $i$, where $x_{-i}\in X_{-i}=X_1\times\ldots\times X_{i-1}\times X_{i+1}\times\ldots\times X_m \subseteq \mathbb{R}^{n-n_i}$. We denote the game by $\Gamma=([m],\{J_i\},\{X_i\})$.

A solution $x^*\in X$ to the game $\Gamma$ is an NE if and only if for every agent $i\in[m]$, we have:
\beqn
J_i(x_i^*,x_{-i}^*)\le J_i(x_i,x_{-i}^*), \quad \forall x_i\in X_i.
\eeqn

When, for every agent $i$, the action set $X_i$ is closed and convex, and the cost function $J_i(x_i,x_{-i})$ is also convex and differentiable in $x_i$ for each $x_{-i}\in X_{-i},$ an NE $x^*\in X$ of the game 
can be alternatively characterized using the first-order optimality conditions. Specifically, 
$x^*\in X$ is an NE of the game if and only if, for all $i\in[m]$:
\beqn
\langle \nabla_{i} J_i(x_i^*,x_{-i}^*),x_i-x_i^*\rangle\ge 0, \quad  \forall x_i \in X_i.
\eeqn

Using the Euclidean projection property yields: 
\beqn
\label{eq-agent-fixed-point}
x_i^*=\Pi_{X_i}[x_i^*-\a_i \nabla_{i} J_i(x_i^*,x_{-i}^*)], \quad \forall i\in[m],
\eeqn
where $\a_i>0$ is an arbitrary scalar.
Stacking the relations in~\eqref{eq-agent-fixed-point}, we can rewrite them in a compact form. Consequently, $x^*$ is an NE for the game $\Gamma$ if and only if:
\beqn
\label{eq-fixed-point}
x^*=\Pi_{X} [x^*-F_{\bda}(x^*) ].
\eeqn
Here, 
$F_{\bda}(\cdot)$ is the scaled gradient mapping of the game, 
\begin{align}\label{eq-gradmapping}
F_{\bda}(x)\triangleq \col\left( (\a_i\nabla_i J_i(x_i,x_{-i}))_{i \in [m]}\right),
\end{align}
with $\nabla_i J_i(x_i,x_{-i})=\nabla_{x_i} J_i(x_i,x_{-i})$ for all $i\in[m]$.

In the absence of constraints on the agents' access to other agents' actions, an NE point can be computed through a simple iterative algorithm \cite{FacchineiPang}. 
Starting with some initial point $x_i^0\in X_i$, each agent $i$ updates its decision at time $k$ as follows: 
\begin{equation}\label{eq-basic-algo}
    x_i^{k+1}
=\Pi_{X_i}[ x_i^k - \a_i \nabla_i J_i(x_i^k,x_{-i}^k)].
\end{equation}
This algorithm is guaranteed to converge to an NE under suitable conditions but it requires that every agent $i$ has access to all other agents' decisions $x_{-i}^k$ at all time $k$.

\subsection{Graph-constrained Agents' Interactions}
This paper focuses on the partial-decision information scenario, where a sequence of directed time-varying communication graphs constrains agents' interactions over time.
At time $k$, agents' interactions are constrained by the directed graph $\bbG_k=([m],\E_k)$, where $[m]$ is the agent set and $\E_k$ is the set of directed links. The directed link $(j,i)\in\E_k$ indicates that agent $i$ can receive information from agent $j$ at time $k$. We assume that every node has a self-loop in each graph $\bbG_k$, so that the neighbor sets $\cNinik$ and $\cNoutik$ contain agent $i$ at all times. Specifically, we use the following assumption.
\begin{assumption}\label{asum-graphs}
Each graph $\bbG_k=([m],\E_k)$ is strongly connected and has a self-loop at every node $i\in[m]$.
\end{assumption}

Intuitively, Assumption~\ref{asum-graphs} implies that each agent can communicate with itself (self-loops) and that the graph is not fragmented into disconnected components, since every node is reachable from every other node (strongly-connected).

\begin{remark}\label{rem-graphs}
Assumption~\ref{asum-graphs} can be relaxed by considering a $B$-strongly-connected graph sequence, where it is assumed that there exists an integer $B\ge 1$ such that the graph with edge set 
$\E^B_k=\bigcup_{i=kB}^{(k+1)B-1}\E_i$
is strongly connected for every $k\ge 0$. 
\end{remark}

Associated with each graph  $\bbG_k$ is the weight matrix $W_k$ that is compliant with the connectivity structure of $\bbG_k$, as follows:
\begin{align}\label{eq-mat-compat}
\begin{cases}
[W_k]_{ij}>0, \quad\text{when} j\in\cNinik,\\
[W_k]_{ij}=0, \quad\text{otherwise.}
\end{cases}
\end{align}
We make the following assumption regarding the matrices $W_k$:
\begin{assumption} \label{asum-row-stochastic}
For each $k\ge0$, the weight matrix $W_k$ is row-stochastic and compatible with the graph $\bbG_k$
i.e., it satisfies relation~\eqref{eq-mat-compat}. 
Moreover, there exists a scalar $w>0$ such that ${\min}^{+}(W_k)\ge w$ for all $k\ge 0$.
\end{assumption}

\begin{remark}
The knowledge of $w$ can be ensured as follows: each agent needs to know the total number $m$ of agents in the system. Without additional graph information, agents can be instructed to use the minimal weight $w=1/m$. If more information is available, such as the maximum in-degree $d$ of the nodes (agents) in any of the communication graphs, then agents can be instructed to use $w=1/(d+1)$.
\end{remark}

\begin{remark}
Every agent $i$ controls the entries in the $i$th row of $W_k$, which does not require any coordination of the weights among the agents. In fact, balancing the weights \cite{Bianchi_2021} 
would require some coordination among the agents, 
which we avoid imposing in this paper.
\end{remark}

\subsection{Partial-Decision Information Scenario}
To address the lack of global knowledge, each agent $i \in [m]$ maintains an estimate vector
$z_i^{k}= \col\left( (z_{ij}^k)_{ j\in [m]}\right)\in\re^n$ where $z_{ij}^k \in \mathbb{R}^{n_j}$ is agent $i$'s estimate of the action $x_j$ of agent $j \neq i$ at time $k$, while $z_{ii}^k=x_i^k$. We define $z_{j,-i}^k= \col\left( (z_{j\ell}^k)_{ \ell\in [m],\ell\neq i}\right)\in\re^{n-n_i}$ as the estimate of agent $j$ without the $i$-th block-component.

Given that the game $\Gamma$ has constraints on agents' access to the actions of other agents, we adapt the basic algorithm~\eqref{eq-basic-algo} to comply with the information access dictated by the graph $\bbG_k$ at time $k$.
In the partial-decision information scenario, agent $i$ will utilize an estimate $v_{i}^{k+1}$, leading to the following update:
\begin{equation*}
x_i^{k+1}
=\Pi_{X_i}\left[ v_{ii}^{k+1}-\a_i \nabla_i J_i\left(v_{i}^{k+1}\right)\right], \quad \forall i\in[m].
\end{equation*}
Here, $v_{i}^{k+1} = \col\left( (v_{ij}^{k+1})_{ j\in [m]}\right)\in\re^n$ is agent $i$'s \textit{aggregated estimate} of the action profile $x$ at time $k+1$. This estimate is a weighted average of its own previous estimate $z_i^k$ and the estimates $z_j^k$ received from its neighbors $j \in\cNinik$, through information exchange dictated by the graph $\bbG_k$.

In this scenario, $v_{ii}^{k+1}$ needs not belong to $X_i$ at any time $k$, as other agents may not know this set. Also, since agent $i$ lacks knowledge of $X_{-i}$, $z_{i,-i}^{k+1}$ and $v_{i,-i}^{k+1}$ need not belong to $X_{-i}$. Thus, $J_i(\cdot)$,  $\forall i\in[m]$, should be defined on $\mathbb{R}^{n}$. Specifically, regarding cost functions and action sets, we assume:


\begin{assumption}\label{assum:lip}
Consider the game $\Gamma$,
assume for all $i \in [m]$:
\begin{itemize}
\item[(a)] The mapping $\nabla_i J_i(x_i,\!\cdot)$ is Lipschitz continuous on $\re^{n-n_i}\!$ for every $x_i \in \re^{n_i}$ with a uniform constant $L_{-i}>0$.
\item[(b)]
 The mapping 
 $\nabla_i J_i(\cdot,x_{-i})$ is 
 Lipschitz continuous on $\re^{n_i}$ for every $x_{-i}\in \re^{n-n_i}$ with a uniform constant $L_i>0$.
 \item[(c)] The set $X_i$ is nonempty, convex, and closed.
 \end{itemize}  
\end{assumption}

\begin{assumption}\label{assum:map_monotone}
The game mapping $F(x):\re^n\to\re^n$,
\begin{align}\label{eq:gamemapping}
F(x)\triangleq \col\left( (\nabla_i J_i(x_i,x_{-i}))_{i \in [m]}\right), 
\end{align}
is strongly monotone with the constant $\mu$.
\end{assumption}

\begin{remark}
\revtwo{The Lipschitz continuity of $\nabla_i J_i(x_i, x_{-i})$ in both $x_i$ and $x_{-i}$ in Assumption~\ref{assum:lip} ensures that $F(x)$ is continuous on $\mathbb{R}^n$.
Assumption~\ref{assum:map_monotone} further implies strong convexity of each cost function $J_i(x_i,x_{-i})$ on $\re^{n_i}$ for every $x_{-i}\in \re^{n-n_i}$ with the constant $\mu$, as noted in Remark 1 of \cite{Tatarenko2021}.
Additionally, under Assumptions~\ref{assum:lip} and~\ref{assum:map_monotone}, an NE point exists and it is unique (Theorem 2.3.3 of \cite{FacchineiPang}). This NE point can also be equivalently represented as a fixed-point solution (see~\eqref{eq-fixed-point}).}
\end{remark}

\begin{remark}
The monotonicity assumption is prevalent in the literature on distributed algorithms for NE seeking. Gradient-based methods typically require strong monotonicity for fixed stepsizes  \cite{Belgioioso2018,Salehisadaghiani_Wei_Pavel_AUT_2019,Tatiana2018,Tatiana2020,Tatarenko2021,Bianchi2020NashES,Bianchi_2021,Huang_Hu_arXiv_2021,Nguyen2023AccGame,GadjovTCSN2023}, strict monotonicity for vanishing stepsizes \cite{Koshal2016,Farzad2019}, or monotonicity with diminishing Tikhonov regularization \cite{Lei_Shanbhag_Chen_Regularization_CDC_2020}. Several practical applications satisfy Assumptions \ref{assum:lip}--\ref{assum:map_monotone}, including Nash-Cournot game \cite{Belgioioso2018,Salehisadaghiani_Wei_Pavel_AUT_2019,Tatiana2018,Tatiana2020,Tatarenko2021,Bianchi2020NashES,Bianchi_2021,Huang_Hu_arXiv_2021,Nguyen2023AccGame,Farzad2019,Lei_Shanbhag_Chen_Regularization_CDC_2020}, mobile sensor networks  \cite{GadjovTCSN2023,Stankovic_TAC_2012}, crowdsourcing \cite{wiopt23} and multi-product assembly problems \cite{Huang_Hu_arXiv_2021}.
\end{remark}



\section{Distributed Algorithm} \label{sec:algo}
\begin{algorithm}[t!]
\caption{Distributed Nash Equilibrium Seeking}
\begin{algorithmic}[1]
\State \textbf{Initialization:} Each agent $i\in[m]$ initializes with $z_{ii}^0\in X_i$ and $z_{i,-i}^0\in\re^{n-n_i}$
\For{$k=0,1,\ldots,$ every agent $i\in[m]$ }
\State  Receives $z_j^k$ from in-neighbors $j\in\cNinik$
\State  Chooses the weights $[W_k]_{ij}$, $\forall j\in[m]$
\State  Sends $z_i^k$ to  out-neighbors $\ell\in\cNoutik$
\State  Performs consensus update: 
\begin{align}\label{eq-viimix}
\textstyle v_{i}^{k+1}=\sum_{j=1}^m [W_k]_{ij}z_{j}^k
\end{align}
\State Updates action:
\begin{align}\label{eq-algo}
x_{i}^{k+1}=\Pi_{X_i}[ {v_{ii}^{k+1}} -\a_i^{k+1}\nabla_i J_i(v_{i}^{k+1})]
\end{align}
\State Updates estimates: $z_{ii}^{k+1} = x_i^{k+1}$, and $z_{i,-i}^{k+1}=v_{i,-i}^{k+1}$
    \EndFor
\end{algorithmic}
\label{ALG:DNE}
\end{algorithm}

We consider the distributed algorithm over a sequence $\{\bbG_k\}$ of directed communication graphs. At time $k$, every agent $i$ sends its estimate $z_i^k$ to its out-neighbors $\ell\in\cNoutik$ and receives estimates $z_j^k$ from its in-neighbors $j\in\cNinik$. 
After exchanging information, agent $i$ performs an average consensus update based on the estimates received from its in-neighbors (see \eqref{eq-viimix}).
Subsequently, agent $i$ updates its own action according to \eqref{eq-algo}. Agent $i$ then updates its estimate of the action profile $z_{i}^{k+1}$, setting $z_{ii}^{k+1} \!=\! x_i^{k+1}$ (its updated action) and $z_{i,-i}^{k+1}\!=\!v_{i,-i}^{k+1}$ (aggregated estimates of other agents' actions). Note that the vectors $z_i^k$ and $v_i^k$, for all $k\ge 0$, are almost identical, except at element $i$ (meaning $z_{ii}^{k} \neq v_{ii}^{k}$ and $z_{i,-i}^{k}=v_{i,-i}^{k}$). The procedure is summarized in Algorithm 1.


\begin{remark}
In the convergence analysis, particularly in relation \eqref{eq-piNormEstFromWeightedAvg}, we demonstrate that under Assumption~\ref{asum-row-stochastic}, the agents' estimates of the action profile $x$ become more closely aligned with the average estimate after the information is mixed using \eqref{eq-viimix}. This motivates substituting $x_i^k$ in \eqref{eq-basic-algo} with $v_{ii}^{k+1}$ in \eqref{eq-algo}.
\end{remark}

\begin{remark}[Initial Value Selection]
The initialization of the estimate vector, including $z_{ii}^0 \in X_i$ and $z_{i,-i}^0 \in \mathbb{R}^{n-n_i}$, can be random while ensuring convergence, as will be shown in Theorem~\ref{theo:convTheo}. While closer initial values can enhance convergence speed, in distributed settings with limited information, the tradeoff is between the benefits of improved convergence and the effort required for more precise initialization.
\end{remark}

\section{Preliminary results}\label{sec:basicre}
This section presents preliminary results on vector norms, graphs, stochastic matrices, and the gradient method.
\vspace{-0.3cm}

	\subsection{Linear Combinations and Graphs}\label{ssec-lincomb-graphs}
    
	
	Since the mixing term $\sum_{j=1}^m\![W_k]_{ij}z_j^k$ used in Algorithm 1 represents a linear combination of $z_j^k$, we begin by deriving a relation for the squared norm of such linear combinations, which will be used in our analysis with various identifications. 
	
	\begin{lemma}\label{lem-normlincomb}
	Let $\{u_i, \, i\in[m]\}\subset\re^n$ be a collection of $m$ vectors and $\{\g_i,\, i\in[m]\}$ 
	be a collection of $m$ scalars. \\
	(a) We have: \[ \displaystyle
	\left\|\sum_{i=1}^m \g_i u_i\right\|^2
	\!\!\!=\!\Bigg(\sum_{j=1}^m \g_j\!\Bigg)\!\!\sum_{i=1}^m  \g_i \|u_i\|^2
	-\frac{1}{2}\sum_{i=1}^m \!\sum_{j=1}^m \g_i \g_j \|u_i-u_j\|^2\!.\]
	(b) If $\sum_{i=1}^m \g_i=1$ holds, then for all $u\in \re^n$ we have:\\
	\[
	\left\|\sum_{i=1}^m \g_i u_i - u \right\|^2
	\!\!\!=\! \sum_{i=1}^m \g_i \|u_i-u\|^2
	-\frac{1}{2}\sum_{i=1}^m \sum_{j=1}^m \g_i \g_j \|u_i-u_j\|^2\!.\]
	\end{lemma}
\begin{proof}
See Appendix~\ref{app:lem-normlincomb}.
\end{proof}

\noindent
The following relations are immediate results of Lemma~\ref{lem-normlincomb}(b):
	
\begin{corollary}\label{cor-averages}
\!\!\! Choosing $u\!=\!\sum_{\ell=1}^m \!\g_\ell u_\ell$ in Lemma~\ref{lem-normlincomb}(b) yields
\begin{equation}\label{eq-aver-disp}
\frac{1}{2}\sum_{i=1}^m \!\sum_{j=1}^m \g_i \g_j \|u_i-u_j\|^2 =\!\sum_{i=1}^m \g_i \left\|u_i -\! \sum_{\ell=1}^m \g_\ell u_\ell\right \|^2\!.
\end{equation}
Substituting \eqref{eq-aver-disp} into Lemma~\ref{lem-normlincomb}(b) we obtain for all $u\in \re^n$:
\begin{equation}\label{eq-gen-aver}
\!\left\|\sum_{i=1}^m \g_i u_i - u \right\|^2 \!\!\!=\!\!\sum_{i=1}^m \g_i \|u_i-u\|^2 \!-\!\!\sum_{i=1}^m \g_i \!\left\|u_i \!-\!\!\sum_{\ell=1}^m \g_\ell u_\ell\right \|^2\!\!\!.\!
\end{equation}
If additionally, $\g_i \ge 0$ for all $i\in [m]$, then \eqref{eq-gen-aver} coincides with the well known relation for weighted averages of vectors.
\end{corollary}

There are certain contraction properties of the distributed method, which are inherited from the use of the mixing term $\sum_{j=1}^m[W_k]_{ij}z_j^k$ and the compliance of $W_k$ with the directed strongly connected graph $\bbG_k$. Lemma~\ref{lem-normlincomb} provides a critical result to capture these properties.
However, Lemma~\ref{lem-normlincomb} alone is not sufficient, since it does not make any use of the structure of the matrix $W_k$ associated with the underlying graph $\bbG_k$.
	
The graph structure is exploited in the forthcoming lemma for a generic graph. The lemma establishes an important lower bound on the quantity $\sum_{(j,\ell)\in\E} \|z_j - z_\ell \|^2$ for a given directed graph $\bbG=([m],{\E})$, where $z_i\in\re^n$ is a vector associated with node $i$. This lower bound will be applied to the graph $\bbG_k$ at time $k$, which provides the second critical step leading us towards the contraction properties of the iterate sequences.
	
\begin{lemma}\label{lem-xdiverse}
	Let $\bbG=([m],{\E})$ be a strongly connected directed graph, where a vector $z_i\in\re^n$ is associated with node $i$ for all $i\in[m]$. 
    We then have:
	\[\sum_{(j,\ell)\in\E} \|z_j - z_\ell \|^2
	\ge \frac{2}{\mathsf{D}(\bbG)\mathsf{K}(\bbG)}\sum_{j=1}^m\sum_{\ell=j+1}^m\|z_j - z_\ell\|^2,\]
	where $\mathsf{D}(\bbG)$ is the diameter of the graph and $\mathsf{K}(\bbG)$ is the maximal edge-utility in the graph (see Definitions~\ref{def-diam} and~\ref{def-edgeut}). 
\end{lemma}

\begin{proof}
See Appendix~\ref{app-lem-xdiverse}.
\end{proof}
\vspace{-0.25cm}
	\subsection{Implications of the Stochastic Nature of Matrix $W_k$}\label{sec-stochmat}
	We provide some basic relations due to the row-stochasticity of the matrices. In the forthcoming lemma, we state some convergence properties of the transition matrices $W_kW_{k-1}\ldots W_t$, which are valid under the assumptions of strong connectivity of the graphs $\bbG_k$ and the graph compatibility of the matrices $W_k$. These properties are known for such a sequence of matrices (see~\cite[Lemma 2]{Nedic2015}).
	
\begin{lemma}\label{lemma-TimeVaryPiVectors}
Let Assumption~\ref{asum-graphs} hold and let $\{W_k\}$ be a matrix sequence satisfying Assumption~\ref{asum-row-stochastic}. For all $k\ge 0$, we have
\begin{enumerate}
\item[(a)] There exists a sequence $\{\pi_k\}$ of stochastic vectors such that 
$\pi_{k+1}'W_k=\pi_k'$. 
\item[(b)] The entries of each $\pi_k$ have a uniform lower bound, i.e.,
\[[\pi_k]_i\ge \tfrac{w^m}{m} \quad \hbox{for all $i\in[m]$ and all $k\ge0$},\]
where $w \le {\min}^{+}(W_k)$ for all $k \ge 0$ (see Assumption~\ref{asum-row-stochastic}).
\end{enumerate}
\end{lemma}

\begin{proof}
See Appendix~\ref{app-lemma-TimeVaryPiVectors}.
\end{proof}
	
The stochastic vector $\pi_k$ in Lemma~\ref{lemma-TimeVaryPiVectors} will be used to define an appropriate Lyapunov function associated with the method. The sequence $\{\pi_k\}$ is an absolute probability sequence~\cite{Seneta}, which has also been used in~\cite{touri2014} to study the ergodicity properties of more involved random matrix sequences.
	
\subsection{Contraction Property of Gradient Method}\label{ss-grad}
In this section, we analyze the contraction property of the gradient mapping $F(\cdot)$ of the game.

\begin{lemma}[\cite{Tatarenko2021}, Lemma 1]\label{lemma:LipschitzNablaJi}
Let  Assumptions~\ref{assum:lip}(a) and \ref{assum:lip}(b) hold. 
For all $x,y \in \re^n$, we have
\begin{align}\label{eq-DelJLipschitz}
\|\nabla_i J_i(x)-\nabla_i J_i(y)\|^2\le \left(L_{-i}^2+L_i^2\right)\|x-y\|^2,
\end{align}
where $L_{-i}$ and $L_i$ are the Lipschitz constants.
\end{lemma}
{

In our analysis, we use a mapping $\bF_{\bda}(\cdot):\re^{m\times n}\to \re^{m\times n}$ to capture the updates for all agents $i\in[m]$ at any time $k$. Given a matrix $\bz\in\re^{m\times n}$, let $z_{i:}$ be the vector in the $i$th row of $\bz$. 
The $i$th row of the matrix 
$\bF_{\bda}(\bz)$ is defined by
\begin{equation}\label{eq-def-bfa}
\!\![\bF_{\bda}(\bz)]_{i:}\!=\!(\zero'_{n_1}\!,\!..., \zero'_{n_{i-1}}\!, \a_i(\nabla_i J_i(z_{i:}))'\!, \zero'_{n_{i+1}}\!,\!...,\!\zero'_{n_m} ).\!\!
\end{equation}

\begin{lemma}\label{lemma-LipschitzbF}
Let  Assumptions~\ref{assum:lip}(a) and \ref{assum:lip}(b) hold. 
Consider the mapping $\bF_{\bda}(\cdot)$ defined by~\eqref{eq-def-bfa}. For any weighted norm defined by a stochastic vector $\pi\!>\!\zero$, we have for $\bx,\by\!\in\!\re^{m\times n}$,
\begin{align} 
    \!\|\bF_{\bda}{(\bx)}\!-\bF_{\bda}{(\by))}\|_\pi^2 \!\le \max\limits_{i\in[m]}\{\a_i^2\!\left(L_{-i}^2\!+ L_i^2\right)\}\|\bx \!-\by\|_\pi^2. \!\! \label{eq:LipschitzbF}
\end{align}
\end{lemma}

\begin{proof}
See Appendix~\ref{app-lemma-LipschitzbF}.
\end{proof}

\section{Convergence Analysis}\label{sec:conv}

\subsection{Contraction Property of Weighted Dispersion}
We begin our analysis by considering generic vectors of the form $r_i=\sum_{j=1}^m W_{ij}z_j,~i\in[m]$, where $W$ is an $m\times m$ row-stochastic matrix and $z_j \in \re^n$ for all $j\in[m]$. 
Noting that each vector $r_i$ is a convex combination of the vectors $z_j,~j\in[m]$, we make use of Lemma~\ref{lem-normlincomb} and Lemma~\ref{lem-xdiverse} to obtain an upper bound of some weighted dispersion of the vectors $r_1,\ldots,r_m$ in terms of a weighted dispersion of the original vectors $z_1,\ldots, z_m$, as seen in the forthcoming lemma. 

\begin{lemma}\label{lemma-lemma6PushPull}
Let $\bbG=([m],\E)$ be a strongly connected directed graph, and let $W$ be an $m\times m$ row-stochastic matrix that is compatible with the graph and has positive diagonal entries, i.e., $W_{ij}>0$ when $j=i$ and $(j,i)\in \E$, and $W_{ij}=0$ otherwise. Also, let $\pi$ be a stochastic vector and let $\phi$ be a nonnegative vector such that 
$\phi'W=\pi'$.

Consider a collection of vectors $z_{1},\ldots,z_{m} \in \re^n$ and consider the vectors 
$r_i$ given by $r_i=\sum_{j=1}^m W_{ij}z_j$ for all $i\in[m]$. Then, for all $u\in \re^n$, we have:
\revtwo{
\begin{align*}
    \sum_{i=1}^m\phi_i&\left\lVert r_i-u\right\rVert^2
    \le \sum_{j=1}^m\pi_j\|z_j-u\|^2 -\eta\sum_{j=1}^m\pi_j\|z_j-\hat{z}_{\pi}\|^2,
\end{align*}
where $\hat{z}_{\pi}=\sum_{i=1}^m \pi_{i}z_i$ and $\eta = \frac{\min(\phi)\left({\min}^{+}(W)\right)^2}{\max^2(\pi)\mathsf{D}(\bbG)\mathsf{K}(\bbG)}$.}
\end{lemma}
	
\begin{proof}
	For any $u\in\re^n$, by the definition of $r_i$, we have:
	\begin{align*}
	&\|r_i - u \|^2 =\Bigg\|\sum_{j=1}^m W_{ij}z_j - u \Bigg\|^2\\
	=& \sum_{j=1}^m W_{ij}\|z_j - u\|^2 -\frac{1}{2} \sum_{j=1}^m \sum_{\ell=1}^m W_{ij} W_{i\ell}\left\|z_j - z_\ell \right\|^2,
	\end{align*}
    where we use Lemma~\ref{lem-normlincomb}(b) with $u_j=z_j$ for all $j$, and $\g=W_{i:}$, where $W_{i:}$ is the $i$th row-vector of the matrix $W$.

    \noindent
	Multiplying the inequality by $\phi_i$ and summing over all $i$ yields:
	\begin{align*}
	\sum_{i=1}^m \phi_i\|r_i - u\|^2 
	&=\sum_{i=1}^m \phi_i\sum_{j=1}^m W_{ij}\|z_j -u\|^2 \\
    &- \frac{1}{2}\sum_{i=1}^m \phi_i \sum_{j=1}^m \sum_{\ell=1}^m W_{ij}W_{i\ell}\|z_j - z_\ell \|^2.
	\end{align*}
 
	In the first term on the right-hand side (RHS) of the preceding inequality, we exchange the order of the summation and use the relation $\phi'W=\pi'$, which yields: 
	\begin{align*}
	\sum_{i=1}^m \phi_i\sum_{j=1}^m W_{ij}\|z_j -u\|^2
	&=\sum_{j=1}^m\left( \sum_{i=1}^m \phi_i W_{ij} \right)\|z_j - u\|^2\\ 
    &=\sum_{j=1}^m \pi_j\|z_j - u\|^2.
	\end{align*}
 
	Hence,
	\begin{align}\label{eq-aver-rel1}
	\sum_{i=1}^m \phi_i\|r_i - u\|^2 
	&=\sum_{j=1}^m \pi_j\|z_j - u\|^2\\
    &- \frac{1}{2}\sum_{i=1}^m \phi_i \sum_{j=1}^m \sum_{\ell=1}^m W_{ij}W_{i\ell}\|z_j - z_\ell \|^2.\nonumber
	\end{align}
	To estimate the last term in~\eqref{eq-aver-rel1},
	we exchange the order of the summation and write:
	\begin{align*}
	&\sum_{i=1}^m \phi_i\sum_{j=1}^m \sum_{\ell=1}^m W_{ij} W_{i\ell}\|z_j - z_\ell \|^2\cr
    =&\sum_{j=1}^m \sum_{\ell=1}^m \|z_j - z_\ell\|^2\left(\sum_{i=1}^m \phi_iW_{ij} W_{i\ell}\right)\cr
	\ge& \sum_{j=1}^m \sum_{\ell\in\mathcal{N}_j^{\rm in} } \|z_j - z_\ell\|^2
	\left(\sum_{i=1}^m \phi_i W_{ij} W_{i\ell}\right).
	\end{align*}
	Since the graph $\bbG$ 
	is strongly connected, every node $j$ must have a nonempty in-neighbor set 
	$\mathcal{N}_j^{\rm in}$. 
	Moreover, by Assumption~\ref{asum-graphs}, we have that $W_{jj}>0$ for every $j\in[m]$
	and $W_{j\ell}>0$ for  all $\ell\in\mathcal{N}_j^{\rm in} $. Hence,
	\begin{align*}
	\sum_{i=1}^m \phi_iW_{ij} W_{i\ell} 
	\ge \phi_j W_{jj} W_{j\ell} 
	\ge \phi_j \left(\min_{ij: W_{ij}>0} W_{ij}\right)^2>0. 
	\end{align*}
	Therefore, using the notation $\min(\phi)=\min_{j\in[m]} \phi_j$ and ${\min}^{+}(W)=\min_{ij: W_{ij}>0} W_{ij}$,
	we have
	\begin{align}\label{eq-lowerb}
	&\sum_{i=1}^m \phi_i\sum_{j=1}^m \sum_{\ell=1}^m W_{ij} W_{i\ell}\|z_j - z_\ell \|^2\cr
    \ge& \min(\phi) \left({\min}^{+}(W)\right)^2
	\sum_{j=1}^m \sum_{\ell\in\mathcal{N}_j^{\rm in} } \|z_j - z_\ell\|^2\cr
	=& \min(\phi) \left({\min}^{+}(W)\right)^2\sum_{(\ell,j)\in\E} \|z_j - z_\ell \|^2.
	\end{align}
	
	Next, we use Lemma~\ref{lem-xdiverse} to bound from below  the sum 
	$\sum_{(\ell,j)\in\E} \|z_j - z_\ell \|^2$. Since $\bbG=([m],{\E})$ is a strongly connected directed graph, by Lemma~\ref{lem-xdiverse} it follows that 
	\[\sum_{(j,\ell)\in\E} \|z_j - z_\ell \|^2
	\ge \frac{2}{\mathsf{D}(\bbG)\mathsf{K}(\bbG)}\sum_{j=1}^m\sum_{\ell=j+1}^m\|z_j - z_\ell\|^2,\]
	where $\mathsf{D}(\bbG)$ is the graph diameter and $\mathsf{K}(\bbG)$ is the maximal edge utility. 
	From the preceding relation  and \eqref{eq-lowerb}, we obtain
	\begin{align}\label{eq-lowerb1}
	&\sum_{i=1}^m \phi_i\sum_{j=1}^m \sum_{\ell=1}^m W_{ij} W_{i\ell}\|z_j - z_\ell \|^2 \cr
    \ge & \frac{2\min(\phi) \left({\min}^{+}(W)\right)^2}{\mathsf{D}(\bbG)\mathsf{K}(\bbG)}
	\sum_{j=1}^m\sum_{\ell=j+1}^m\|z_j - z_\ell\|^2.
	\end{align}	
	To get to the $\pi$-weighted average of the vectors $z_j$, we write
	\begin{align*}
	&\sum_{j=1}^m \sum_{\ell=j+1}^m \|z_j - z_\ell \|^2
	=\frac{1}{2}\sum_{j=1}^m \sum_{\ell=1}^m \|z_j - z_\ell \|^2 \\
    \ge & \frac{1}{2\max^2(\pi)}\sum_{j=1}^m \sum_{\ell=1}^m \pi_j \pi_\ell  \|z_j - z_\ell \|^2.
	\end{align*}
	Finally, by using the weighted average dispersion relation~\eqref{eq-aver-disp},
	with $\g_i=\pi_i$  and $u_i=x_i$ for all $i$, we have
	 \[\frac{1}{2}\sum_{j=1}^m \sum_{\ell=1}^m \pi_j \pi_\ell  \|z_j - z_\ell \|^2
	= \sum_{j=1}^m \pi_j \|z_j - \hat z_\pi \|^2,\]
	with $\hat z_\pi =\sum_{\ell=1}^m \pi_\ell z_\ell$.
	Using the preceding two relations and relation~\eqref{eq-lowerb1}, we observe
    \begin{align}\label{eq-lowerb2}
    \sum_{i=1}^m\! \phi_i\!\sum_{j=1}^m \sum_{\ell=1}^m W_{ij} W_{i\ell}\|z_j - z_\ell \|^2
    \ge  \eta
    \sum_{j=1}^m \pi_j \|z_j - \hat z_\pi \|^2\!.\!\!\!
	\end{align}
	By substituting the estimate~\eqref{eq-lowerb2} into relation~\eqref{eq-aver-rel1}, we obtain 
	the desired relation.
\end{proof}

\revtwo{
\begin{remark} \label{remark-lemma6static}
Let $\pi>\zero$ and $\phi>\zero$ in Lemma~\ref{lemma-lemma6PushPull}. By applying the lemma with $r_i=\sum_{j=1}^m W_{ij}z_j$ and an arbitrary vector $u\in\re^n$, the resulting inequality can be compactly written as:
\begin{equation*}
\|W\bz -\one_m u'\|_{\phi}^2
     \le \|\bz - \one_m u'\|_{\pi}^2 - \eta \|\bz-\one_m\hat{z}'_\pi\|_{\pi}^2,
\end{equation*}
where $\bz$ is an $m \!\times\! n$ matrix with rows $z_i'$ and $\hat z_{\pi} \!=\! \sum_{i=1}^m\pi_i z_i$. 
\end{remark}}

\revtwo{
\begin{corollary}\label{corollary-lemma6time}
Applying Lemma~\ref{lemma-lemma6PushPull} to the time-varying matrix $W_k$ associated with the graph $\bbG_k$, and stochastic vectors $\pi_k>\zero$ satisfying $\pi_{k+1}'W_k=\pi'_k$ yields,
for all $u\in\re^n$,
\begin{equation}\label{eq-piNormEstFromAny}
\!\|W_k\bz^k \!\!-\one_m u'\|_{\pi_{k+1}}^2
\!\!\le\! \|\bz^k \!\!-\! \one_m u'\|_{\pi_k}^2 \!\!-\! \eta_k \|\bz^k \!\!-\!\one_m\hat{z}'_{\pi_k}\|_{\pi_k}^2.\!
\end{equation}
where 
$\hat z_{\pi_k}=\sum_{i=1}^m[\pi_k]_i z_{i}^k$.
The parameter $\eta_k$ is given by 
\[\eta_k = \frac{\min(\pi_{k+1})w^2}{\max^2(\pi_k)\mathsf{D}(\bbG_k)\mathsf{K}(\bbG_k)}\in (0,1).\] 
\end{corollary}
\begin{proof}
See Appendix~\ref{app-lemma6time}.
\end{proof}
}

\begin{remark}\label{remark-estimateEta}
To handle the time-varying graphs $\bbG_k$ and their associated matrices $W_k$, we will employ Lemma~\ref{lemma-lemma6PushPull} with the absolute probability sequence $\{\pi_k\}$ from Lemma~\ref{lemma-TimeVaryPiVectors}, as indicated by Remark~\ref{corollary-lemma6time}. In doing so, we will use a lower bound on the constants $\eta_k$, defined as $\bdeta=\min_{k\ge 0}\eta_k$.
We note  that $\bdeta\in(0,1)$ holds under Assumption~\ref{asum-graphs} and  Assumption~\ref{asum-row-stochastic}. 

Generally, the worst bounds for  $\mathsf{D}(\bbG_k)$ and $\mathsf{K}(\bbG_k)$ are
\[\mathsf{D}(\bbG_k)\le m-1, \qquad\mathsf{K}(\bbG_k)\le m-1\qquad\hbox{for all }k\ge0.\]
By Lemma~\ref{lemma-TimeVaryPiVectors}, the smallest positive entry of $\pi_{k+1}$ is uniformly bounded by $\frac{w^m}{m}$, while $\max(\pi_k)\!\!\le\! 1$. This yields the following extremely pessimistic lower bound for $\eta_k$:
\[\eta_k\ge \tfrac{w^{m+2}}{m(m-1)^2}\qquad\hbox{for all }k\ge0.\]
If the graphs $\bbG_k$ are more structured, a tighter bound can be obtained. For example, if each graph $\bbG_k$ is a directed cycle and has self-loops at every node, then 
\[\mathsf{D}(\bbG_k)= m-1, \qquad \mathsf{K}(\bbG_k)= \tfrac{m(m-1)}{2} \qquad \hbox{for all } k\ge 0.\]
However, in this case, each graph $\bbG_k$ is $2$-regular, and if the diagonal entries of $W_k$ are all equal to $1-w$, with $w\in(0,1)$, then we would have $\pi_k=\frac{1}{m}\ones$ for all $k$, yielding:
\[\eta_k\ge \tfrac{2 w^2}{(m-1)^2} \qquad\hbox{for all }k\ge0.\]
By choosing $w=\frac{m-1}{m}$, we would obtain
$\eta_k\ge \frac{2}{m^2}$ for all $k\ge0$.
While exploring the graph structures that yield tighter lower bounds for $\eta_k$ is interesting on its own, it is not further addressed in this paper. 
For the remainder, we simply use the fact that a lower bound $\bdeta\in(0,1)$ for $\eta_k$, $k\ge0$, exists.
\end{remark}

\subsection{Convergence of Algorithm~1}\label{sec:ConvergenceAlgorithm1}
Consider Algorithm~1, and let $z_i^{k}= \col\left( (z_{ij}^k)_{ j\in [m]}\right)\in\re^n$ for all $k\ge0$. 
Let $\{\pi_k\}$ be the sequence of stochastic vectors satisfying $\pi_{k+1}'W_{k}=\pi_k'$ with $\pi_k>\zero$
for all $k$. 
We define
\begin{equation}\label{eq-notat}
    \bz^{k}=[z_1^{k},\ldots,z_m^{k}]', ~~
    \hat{\bz}_{\pi_{k}}=\one_m(\hat{z}_{\pi_k})',~~
    \bx^*=\one_m(x^*)',
\end{equation}
where $\hat{z}_{\pi_k}=\sum_{i=1}^m[\pi_k]_i z_i^k$ and $x^*$ is the NE of the game. 
We denote $\bda^k=\col\left( (\a_{i}^k)_{ i\in [m]}\right)$, $\bar{\a} =\max_k\{ \max(\bda^k)\}$ and $\underline{\a} = \min_k\{\min(\bda^k)\}$. We denote $\tilde{\a}_{\pi_{k}} = \frac{1}{m}\sum_{i=1}^m [\pi_{k}]_i\a_i^k$, $\tilde{\bda}_{\pi_{k}}=\tilde{\a}_{\pi_{k}}\one_m$, $\epsilon_{\a}^{k} = \|\Diag(\pi_{k})\bda^k -\tilde{\bda}_{\pi_{k}}\|/\tilde{\a}_{\pi_{k}}$ and $\bar{\epsilon}_{\a}=\max_{k}\epsilon_{\a}^{k}$. We further define the following constants:
\begin{align}\label{eq-constants}
\!L \!=\! \sqrt{\max_{i\in[m]}\left(L_{-i}^2+ L_i^2\right)}, ~\bL_{\a}\!=\bar{\a}L, ~\bdbeta_{\a}\!=(\mu - L\bar{\epsilon}_{\a})\underline{\a}.\!
\end{align}

Our next result provides a basic relation for 
the time-varying $\pi_k$-weighted norm of the difference between $\bz^{k}$ and $\bx^*$.

\begin{lemma}\label{lem-basic}
Let Assumption~\ref{asum-graphs}, Assumption~\ref{asum-row-stochastic}, Assumption~\ref{assum:lip}, and Assumption~\ref{assum:map_monotone} hold. Consider Algorithm~1, the notations in~\eqref{eq-notat} and \eqref{eq-constants}. We have for all $k \geq 0$,
\begin{align*}
    \|\bz^{k+1}\!\!-\!\bx^*\|_{\pi_{k+1}}^2 
    \!\!&\le\!\! \left(\!1\!+\!\bL_{\a} ^2\right)\!\!\|W_k\bz^k \!\!-\!\bx^*\|_{\pi_{k+1}}^2 \!\!\!-\!2\bdbeta_{\a}\|\hat{\bz}_{\pi_{k}}\!\!-\! \bx^*\|_{{\pi_{k}}}^2\nonumber\\
    &+2\bL_{\a} \|W_k\bz^k-\bx^*\|_{\pi_{k+1}}\|W_k\bz^k-\hat{\bz}_{\pi_{k}}\|_{\pi_{k+1}}\\
    &+2\bL_{\a} \|W_k\bz^k\!-\!\hat{\bz}_{\pi_{k}}\|_{\pi_{k+1}}\|\hat{\bz}_{\pi_{k}}\!-\!\bx^*\|_{\pi_{k+1}}.
    \end{align*}
\end{lemma}
\begin{proof}
See Appendix~\ref{app-lem-basic}.
\end{proof}


We then provide the convergence result for Algorithm~1.
\begin{theorem} \label{theo:convTheo}
Let Assumption~\ref{asum-graphs}, Assumption~\ref{asum-row-stochastic}, Assumption~\ref{assum:lip}, and Assumption~\ref{assum:map_monotone} hold. Consider Algorithm~1 and the notations in~\eqref{eq-notat}--\eqref{eq-constants}.
For all $k \geq 0$, the following holds:
\begin{align*}
\|\bz^{k+1}-\bx^*\|_{\pi_{k+1}}^2 \le \lmax{\bar{Q}_{\bda}}\| \bz^k-\bx^*\|_{\pi_{k}}^2,\
\end{align*}
where the matrix $\bar{Q}_{\bda}$ is given by
\[\bar{Q}_{\bda}=\begin{bmatrix}
1-2\bdbeta_{\a}+\bL_{\a} ^2 & 2\sqrt{1- \bdeta}\bL_{\a} \\
2\sqrt{1- \bdeta}\bL_{\a}  & (1+2\bL_{\a} +\bL_{\a} ^2)(1- \bdeta)
\end{bmatrix}.\]
If the stepsizes are chosen such that  $\lmax{\bar{Q}_{\bda}}\!<\!1$, 
then 
\begin{align*}
\lim_{k\to\infty}\|\bz^k-\bx^*\|=0,\qquad \lim_{k\to\infty}\|x^k-x^*\|=0.
\end{align*}
\end{theorem}

\begin{proof}
The definitions of $\hat{\bz}_{\pi_{k}}$ and $\bx^*$ in~\eqref{eq-notat} imply that
\beqn \label{eq:EqPiTime}
    \|\hat{\bz}_{\pi_{k}}-\bx^*\|_{\pi_{k+1}}=\|\hat{\bz}_{\pi_{k}}-\bx^*\|_{\pi_k}.
\eeqn

Substituting \eqref{eq:EqPiTime} into Lemma~\ref{lem-basic}, we have 
for all $k\ge0$,
\begin{align} \label{eq-piNormEstFromWeightedAvgk}
    \|\bz^{k+1}\!\!&-\!\bx^*\|_{\pi_{k+1}}^2 
    \!\!\le\!\! \left(\!1\!+\!\bL_{\a} ^2\right)\!\!\|W_k\bz^k \!\!-\!\bx^*\|_{\pi_{k+1}}^2 \!\!\!-\!2\bdbeta_{\a}\|\hat{\bz}_{\pi_{k}}\!\!-\! \bx^*\|_{{\pi_{k}}}^2\nonumber\\
    &+2\bL_{\a} \|W_k\bz^k-\bx^*\|_{\pi_{k+1}}\|W_k\bz^k-\hat{\bz}_{\pi_{k}}\|_{\pi_{k+1}}\nonumber\\
    &+2\bL_{\a} \|W_k\bz^k\!-\!\hat{\bz}_{\pi_{k}}\|_{\pi_{k+1}}\|\hat{\bz}_{\pi_{k}}\!-\!\bx^*\|_{\pi_{k}}.
\end{align}

The main idea of the remainder of the proof is to establish the evolution relation for the quantity
$\|\bz^{k+1}-\bx^*\|_{\pi_{k+1}}$ in terms of $\|\bz^k-\hat{\bz}_{\pi_{k}}\|_{\pi_k}$ and $\|\hat{\bz}_{\pi_{k}}-\bx^*\|_{\pi_k}$.
To achieve this, we apply Corollary~\ref{corollary-lemma6time} with $u = x^*$, yielding:
\begin{equation}\label{eq-piNormEstFromNash}
\|W_k\bz^k -\bx^*\|_{\pi_{k+1}}^2
     \le \|\bz^k - \bx^*\|_{\pi_k}^2 - \eta_k \|\bz^k-\hat{\bz}_{\pi_{k}}\|_{\pi_k}^2.
\end{equation}
Using relation~\eqref{eq-piNormEstFromAny} in Corollary~\ref{corollary-lemma6time} with
$u=\hat{z}_{\pi_k}$, we obtain:
\begin{equation}\label{eq-piNormEstFromWeightedAvg}
\|W_k\bz^k - \hat{\bz}_{\pi_{k}}\|_{\pi_{k+1}}^2
     \le (1-\bdeta) \|\bz^k-\hat{\bz}_{\pi_{k}}\|_{\pi_k}^2.
\end{equation} 
Finally, using Corollary~\ref{cor-averages} with $\g_i=[\pi_k]_i$ and $u_i=z_i^k$, we observe that for any $u\in\re^n$,
\begin{equation*}
\|\hat{z}_{\pi_{k}}- u \|^2 =\|\bz^k-\ones_m u'\|^2_{\pi_k}
         -\|\bz^k - \hat{\bz}_{\pi_{k}}\|^2_{\pi_k}.
\end{equation*}
Letting $u=x^*$ and noting  
$\|\hat{z}_{\pi_{k}} \!- x^* \|^2 \!=\|\hat{\bz}_{\pi_{k}} \!- \bx^*\|^2_{\pi_k}$,
yields:
\begin{equation}\label{eq-zaverandNash}
\|\bz^k-\bx^*\|^2_{\pi_k}
=\|\bz^k - \hat{\bz}_{\pi_{k}}\|^2_{\pi_k}
+\|\hat{\bz}_{\pi_{k}}- \bx^* \|^2_{\pi_k}.
\end{equation}
Using relation \eqref{eq-zaverandNash} for the first term in \eqref{eq-piNormEstFromNash}, we obtain:
\begin{equation}\label{eq-piNormEstFromNash1}
\|W_k\bz^k \!\!-\!\bx^*\|_{\pi_{k+1}}^2
\!\!\le \|\hat{\bz}_{\pi_{k}} \!\!-\! \bx^* \|^2_{\pi_k} \!\!+\!(1 \!-\! \bdeta) \|\bz^k \!\!-\!\hat{\bz}_{\pi_{k}}\|_{\pi_k}^2.
\end{equation}
Moreover, by the triangle inequality, \eqref{eq:EqPiTime} and \eqref{eq-piNormEstFromWeightedAvg}, we have
\begin{align}\label{eq-piNormEstFromNash2}
&\|W_k\bz^k -\bx^*\|_{\pi_{k+1}} \le \|W_k\bz^k -\hat{\bz}_{\pi_{k}}\|_{\pi_{k+1}} + \|\hat{\bz}_{\pi_{k}} -\bx^*\|_{\pi_{k+1}} \nonumber\\
&\le  \sqrt{1- \bdeta} \|\bz^k-\hat{\bz}_{\pi_{k}}\|_{\pi_k} + \|\hat{\bz}_{\pi_{k}}- \bx^* \|_{\pi_k}.
\end{align}

Substituting relations \eqref{eq-piNormEstFromWeightedAvg}, \eqref{eq-piNormEstFromNash1}, and \eqref{eq-piNormEstFromNash2} into \eqref{eq-piNormEstFromWeightedAvgk}, yields:
\begin{align*}
    &\|\bz^{k+1}-\bx^*\|_{\pi_{k+1}}^2\\
    \le&\left(1+\bL_{\a} ^2\right)((1- \bdeta)\| \bz^k-\hat{\bz}_{\pi_{k}}\|_{\pi_{k}}^2+\|\hat{\bz}_{\pi_{k}}-\bx^*\|_{\pi_{k}}^2)\nonumber\\
    +&2\bL_{\a} (\!\sqrt{1\!-\! \bdeta}\| \bz^k\!\!-\!\hat{\bz}_{\pi_{k}}\!\|_{\pi_{k}}\!\!+\!\|\hat{\bz}_{\pi_{k}}\!\!-\!\bx^*\!\|_{\pi_{k}})\!\sqrt{1\!-\! \bdeta}\|\bz^k\!\!-\!\hat{\bz}_{\pi_{k}}\!\|_{\pi_{k}}\nonumber\\
    +&2\bL_{\a}  \sqrt{1\!-\! \bdeta}\|\bz^k\!-\!\hat{\bz}_{\pi_{k}}\|_{\pi_{k}}\|\hat{\bz}_{\pi_{k}} \!-\! \bx^*\|_{\pi_{k}} -2\bdbeta_{\a}\|\hat{\bz}_{\pi_{k}}\!-\!\bx^*\|_{{\pi_{k}}}^2 \nonumber\\
    =& \begin{bmatrix}
    \|\hat{\bz}_{\pi_{k}}-\bx^*\|_{\pi_{k}}\\ 
    \| \bz^k-\hat{\bz}_{\pi_{k}}\|_{\pi_{k}}
    \end{bmatrix}' \bar{Q}_{\bda} \begin{bmatrix}
    \|\hat{\bz}_{\pi_{k}}-\bx^*\|_{\pi_{k}}\\ 
    \| \bz^k-\hat{\bz}_{\pi_{k}}\|_{\pi_{k}}
    \end{bmatrix}.
    \end{align*} 
Hence,
\begin{align*}
\|\bz^{k+1}\!-\!\bx^*\|_{\pi_{k+1}}^2
&\le \lmax{\bar{Q}_{\bda}} (\|\hat{\bz}_{\pi_{k}} \!-\! \bx^*\|_{\pi_{k}}^2 \!\!+\! \| \bz^k \!-\!\hat{\bz}_{\pi_{k}}\|_{\pi_{k}}^2)\\
&=\lmax{\bar{Q}_{\bda}}\| \bz^k-\bx^*\|_{\pi_{k}}^2, 
 \end{align*} 
where the last equality is derived from \eqref{eq-zaverandNash}. The rest of the statement follows immediately from the previous relation.
\end{proof}

Theorem~\ref{theo:convTheo} establishes that the iterates $\{x^k\}$ and the estimates $\{z_i^k\}$ (for all agents $i \in [m]$) generated by Algorithm~1 converge to the NE $x^*$ at a geometric rate \revtwo{in the weighted norm $\|\cdot\|_{\pi_k}$}, provided the stepsizes are appropriately chosen.
 



\begin{remark}\label{remark-Sylvester}
$\lmax{\bar{Q}_{\bda}}<1$ if and only if the matrices $\bar{Q}_{\bda}$ and $I-\bar{Q}_{\bda}$ are positive definite. According to Sylvester's criterion, the following inequalities should hold:
$[\bar{Q}_{\bda}]_{1,1}>0$, $\det(\bar{Q}_{\bda})>0$, $[I-\bar{Q}_{\bda}]_{1,1}>0$, and  $\det(I-\bar{Q}_{\bda})>0$. These conditions can be satisfied by choosing appropriate stepsizes, and an explicit upper bound can be obtained as presented in \cite[Lemma 2]{Bianchi_2021}, noting that $\bdbeta_{\a}>0$ when the stepsizes satisfy $\bar{\epsilon}_{\a} < \frac{\mu}{L}$. The constants related to the communication graph in $\bar{Q}_{\bda}$ can be estimated using Lemma~\ref{lemma-TimeVaryPiVectors} and by exploring the graph structures as discussed in  Remark~\ref{remark-estimateEta}. 
It is worth noting that while theoretical bounds offer convergence guarantees and serve as foundational guidelines for stepsize selection, they can be conservative \cite{pshi21} and often require global information, which is impractical in distributed settings. In practice, empirical tuning and adaptive techniques are commonly used to optimize performance \cite{YANG2020_hyperparameter}.
\end{remark} 

\begin{remark}\label{rem-generality}
Algorithm 1 and its convergence analysis generalizes many existing results \cite{Tatiana2020,Bianchi_2021,Bianchi2020NashES}. Firstly, Theorem~\ref{theo:convTheo} applies to the case where the communication network is directed and static \cite{Bianchi2020NashES}, with $W_k=W$ for all $k \ge 1$. When the underlying graph is strongly connected and $W$ is compliant with the graph, it is well-known that $W$ has a unique stochastic left-eigenvector $\pi>\zero$, associated with the simple eigenvalue $\one$, i.e., $\pi'W=\pi'$. In this case, the convergence result for Algorithm~1 follows immediately from Theorem~\ref{theo:convTheo}, with $\pi_k=\pi$ for all $k \ge 1$. Additionally, by choosing stepsizes as in \cite[Alg. 1]{Bianchi2020NashES}, i.e., $\a_i^k = \frac{\a}{\pi_i}$ for all $i\in [m]$ and $k \ge 1$, we ensure $\bar{\epsilon}_{\a} = 0 < \frac{\mu}{L}$ and hence $\bdbeta_{\a}>0$ (see Remark~\ref{remark-Sylvester}), thereby recovering the result presented in \cite{Bianchi2020NashES}. 

Furthermore, Theorem~\ref{theo:convTheo} also addresses the scenario where the communication network is either a static undirected graph \cite{Tatiana2020} or a sequence of time-varying undirected graphs \cite{Bianchi_2021}. Here, the matrix $W_k$ is doubly stochastic for all $k$, and the left eigenvector associated with the eigenvalue $\one$ is $\pi_k = \frac{1}{m}\one_m$. Thus, by choosing identical constant stepsizes $\a_i^k = \a$ for all $i\in [m]$ and $k \ge 1$, then $\bar{\epsilon}_{\a} = 0 < \frac{\mu}{L}$, thereby recovering the convergence result presented in \cite[Theorem 1]{Bianchi_2021}. The values of $\a$ ensuring $\lmax{\bar{Q}_{\bda}}<1$ can be determined using the conditions given in Remark~\ref{remark-Sylvester}, as detailed in \cite[Lemma 2]{Bianchi_2021}.
\end{remark}

\begin{remark}
Theorem~\ref{theo:convTheo} uses the assumption that the sequence $\{G_k\}$ consists of strongly connected graphs.  
When this assumption is relaxed (see Remark~\ref{rem-graphs}), the analysis can make use of 
Theorem 4.20 in~\cite{Seneta} stating that there exist a set of absolute probability vectors $\{\pi_k\}$, for all $k\ge 0$, such that $\pi'_{k+r}\left(W_{k+r-1}\ldots W_{k+1}W_{k}\right)=\pi'_k.$
Also, Lemma~\ref{lemma-TimeVaryPiVectors}(b) can be extended to show that
$[\pi_k]_i\ge\tfrac{w^{mB}}{m}$ for all $i\in[m]$ and $k\ge0$.
Using these results, the convergence analysis of Algorithm~1 follosw similarly to our analysis for strongly connected graphs.
\end{remark}

\section{Numerical results}\label{sec:simulation}
In this section, we evaluate the performance of the distributed algorithm through a Nash-Cournot game, as described in \cite{Pavel2020}, over several types of communication networks. Specifically, consider a set of $m$ firms (i.e., agents) involved in the production of a homogeneous commodity. The firms compete over $N$ markets, $M_1,\ldots,M_N$. Denote the market index by $h$, where $h \in [N]$.
Firm $i\in [m]$, participates in the competition in $n_i$ markets, where $n_i\le N$ is a non-negative integer number, by deciding the amount of commodity  $x_i \in \Omega_i =[\zero,\vec{C}_i]$, where 
$\vec{C}_i\in \re^{n_i}$, to be produced and delivered.
A local matrix $B_i\in\re^{N\times n_i}$
indicating which markets firm $i$ participates in:
$$[B_i]_{hj}=\begin{cases}
1, \quad \text{if agent i delivers $[x_i]_j $ to $M_h$,}\\
0, \quad \text{otherwise.}
\end{cases}$$  
Let $n=\sum_{i=1}^m n_i$, $x=[x_i]_{i\in [m]}\in\re^n$, and $B=[B_1,\ldots,B_m]\in\re^{N\times n}$. Then, given an action profile $x$ of all firms, the vector of the total product supplied to the markets can be expressed as $Bx=\sum_{i=1}^mB_ix_i\in \re^N$.

\begin{figure}[t!]
\centering
\includegraphics[scale=0.35]{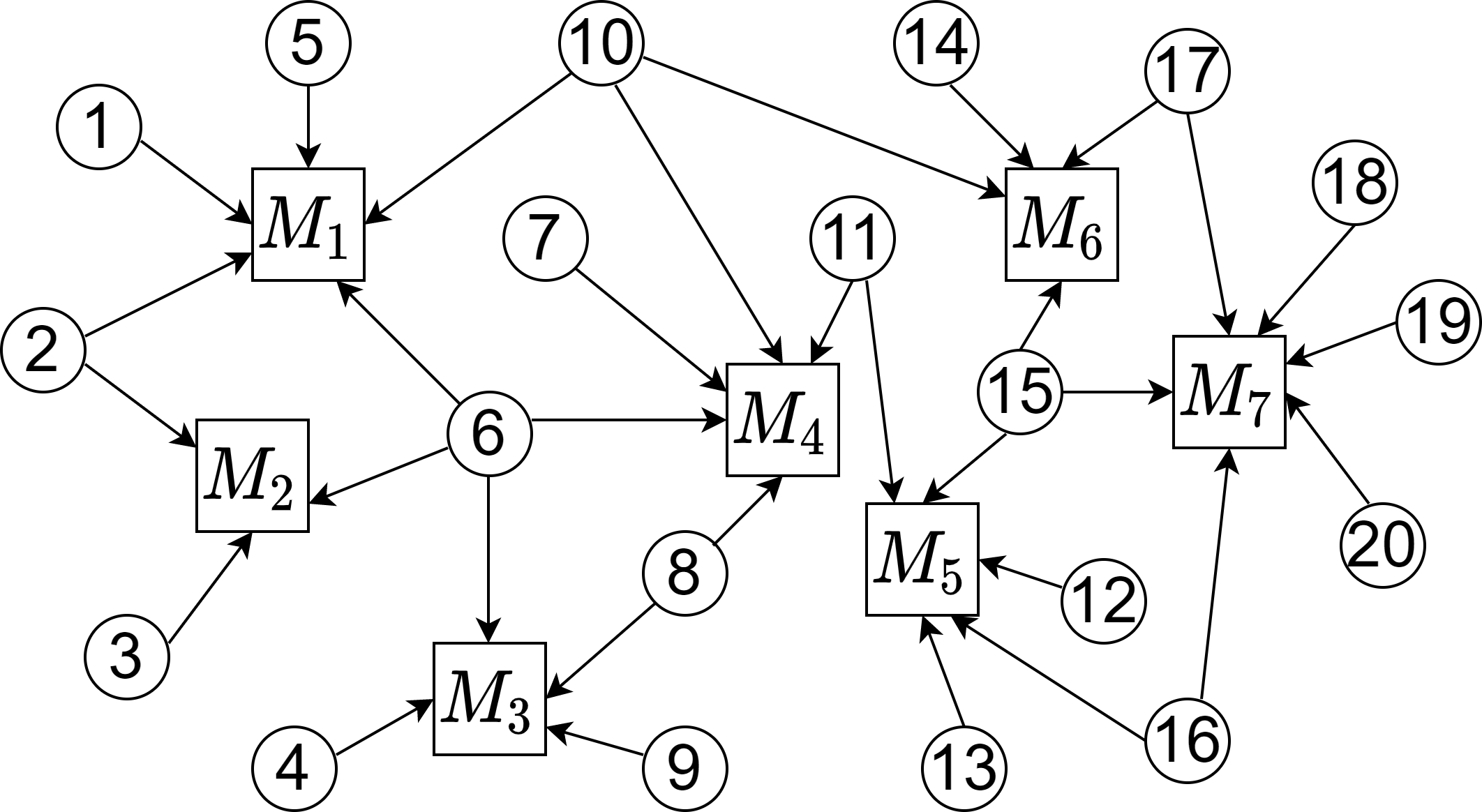}
\caption{Network Nash-Cournot game: An edge from $i$ to $M_h$ on this graph implies that firm $i$ participates in market $M_h$.}\label{fig:NashCournotPavel}
\vspace{-0.6cm}
\end{figure}

\begin{figure}
	\centering
\subfigure[Ring]{
\includegraphics[width=0.142\textwidth]{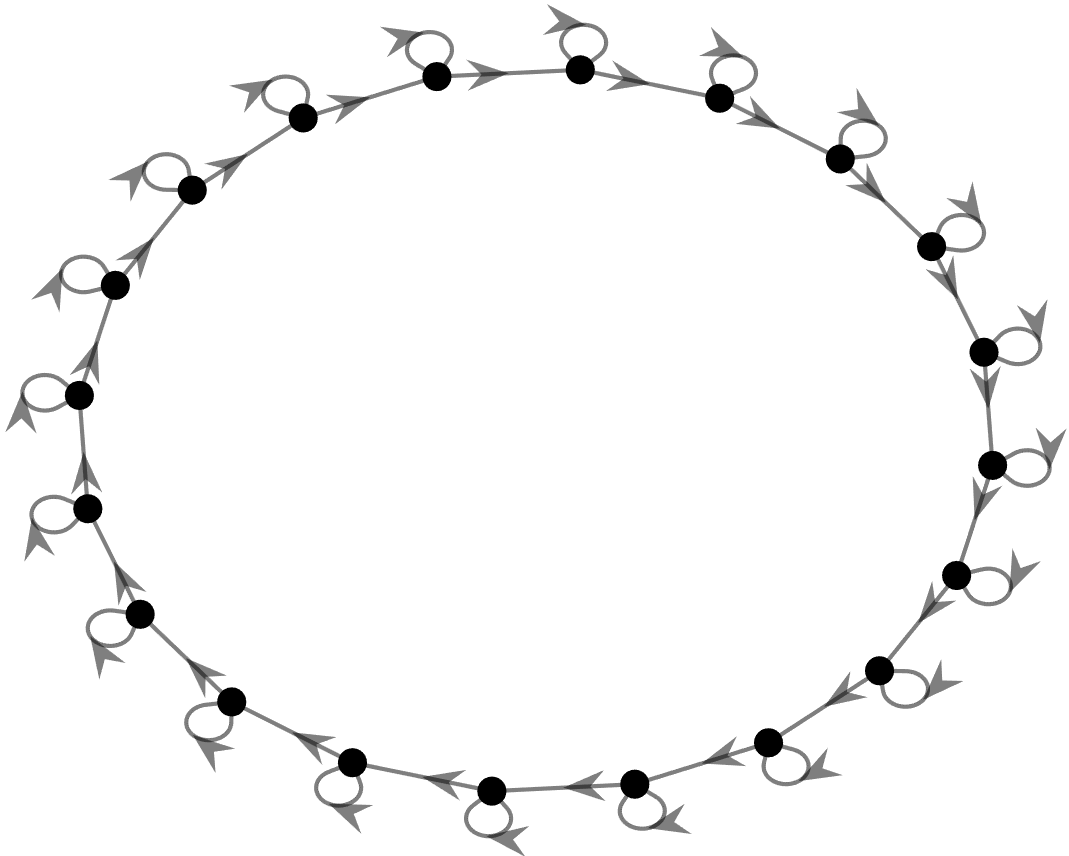}
\label{fig:ring}  } 
\subfigure[Star]{
\includegraphics[width=0.142\textwidth]{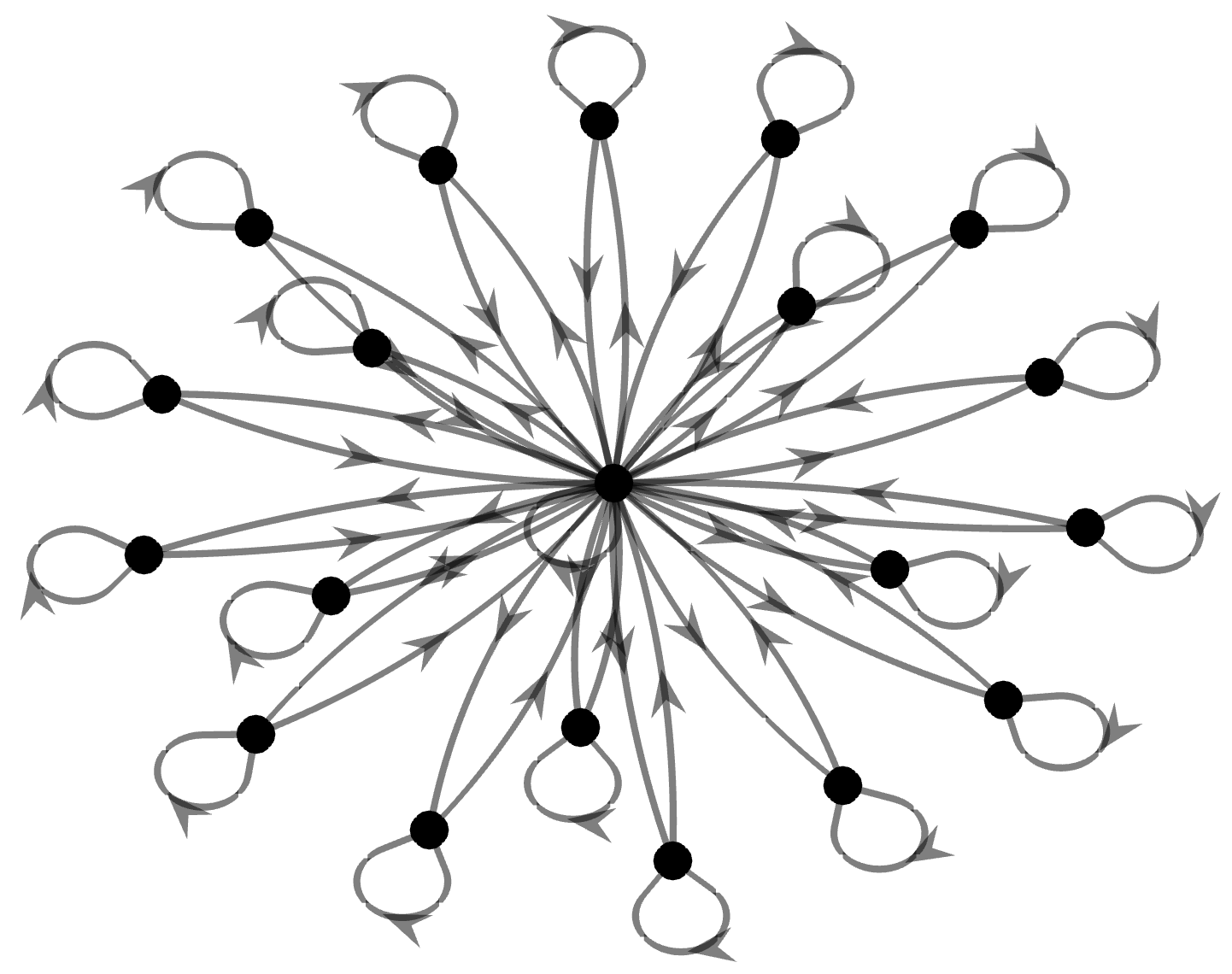}
\label{fig:star}  }
\subfigure[Random]{
\includegraphics[width=0.142\textwidth]{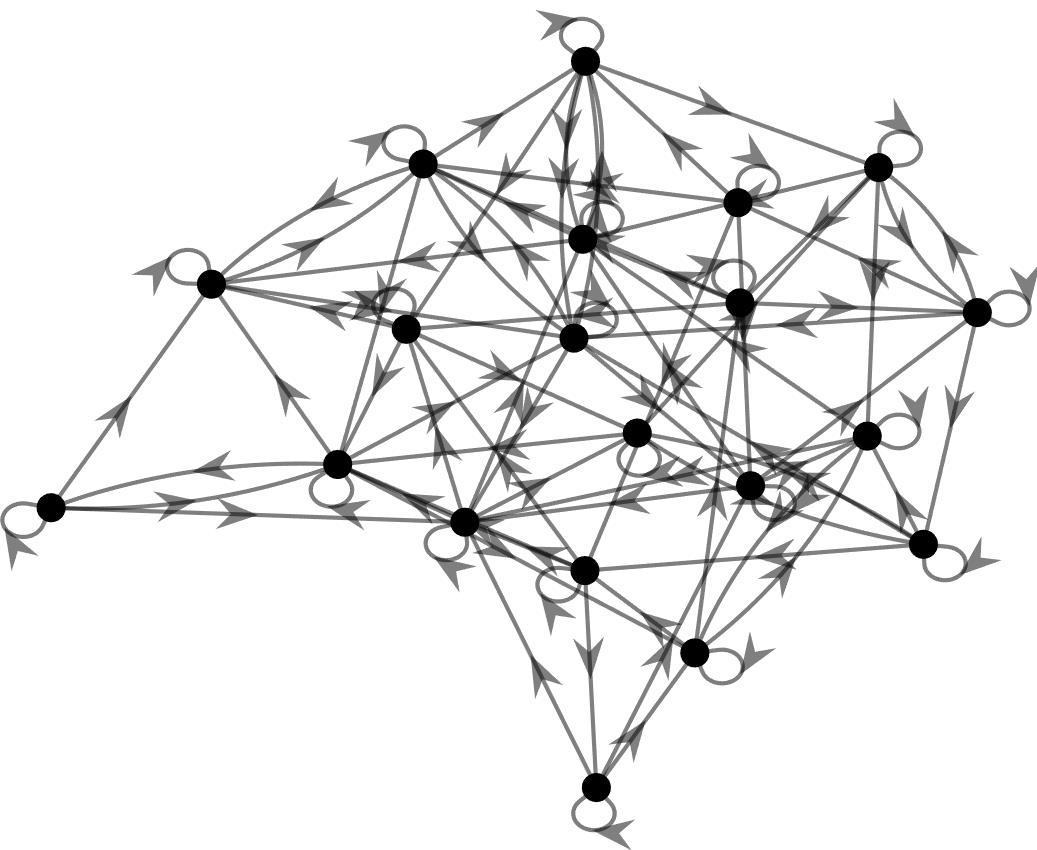}
\label{fig:random}
}  \vspace{-0.2cm}
\caption{Communication network topologies.}
\label{fig:Communication graph}
\vspace{-0.2cm}
\end{figure} 

Similar to \cite{Pavel2020}, we consider a setting with  $m=20$ firms, $N=7$ markets and $n=32$, as presented in Fig.~\ref{fig:NashCournotPavel}. The commodity's price in market $M_h$ is given by:
$p_h(x)=\Bar{P}_h-\chi_h[Bx]_h,~\forall h,$
where $\Bar{P}_h>0$ and $\chi_h>0$. This function  implies that the price decreases as the amount of supplied commodity  increases. Let $\Bar{P}=[\Bar{P}_h]_{h=\overline{1,N}}\in\re^N$ and $\Xi = \diag([\chi_h]_{h=\overline{1,N}})\in\re^{N\times N}$. Then, the price vector function $P=[p_h]_{h=\overline{1,N}}$ that maps the total supply of each market to the corresponding price, has the form: $P=\Bar{P}-\Xi Bx$, and $P'B_ix_i$ is the return of firm $i$ obtained by selling $x_i$  to the markets that it connects with.
Firm $i$’s production cost $c_i(\cdot):\Omega_i \mapsto \re$ is a strongly convex, quadratic function: $c_i(x_i)=x_i'Q_ix_i+q_i'x_i,$
with $Q_i\in \re^{n_i\times n_i}$ symmetric and $Q_i\succ 0$, and $q_i\in\re^{n_i}$.

The local objective function of firm $i$, which depends on the other firms' production profile $x_{-i}$, is given as:
\begin{align}
\label{eq:Ji}
    J_i(x_i,x_{-i})=c_i(x_i)-(\Bar{P}-\Xi Bx)'B_ix_i, ~~ \forall i,
\end{align}
and, $\nabla_i J_i(x)=2Q_ix_i+q_i+B_i'\Xi B_ix_i-B_i'(\Bar{P}-\Xi Bx)$. 

We uniformly sample each entry of $\vec{C}_i$ from the interval $[5,10]$. The matrix $Q_i$ is diagonal with its entries chosen from $[5,8]$, and each $q_i$ is randomly drawn from $[1,2]$. We set $\bar{P}_h$ uniformly within $[10,20]$ and $\chi_h$ within $[0.01, 0.02]$. We consider a partial-decision information scenario and firms may communicate with a local subset of neighbors via a directed network $\bbG_k$ at time $k$. Fig.~\ref{fig:Communication graph} depicts the communication topologies used in our experiments. 

We define the row-stochastic weight matrix $W_k$ 
such that:
$$[W_k]_{ij}=\begin{cases}
0, &\text{if } j\not\in\cNinik,\\
\zeta, &\text{if } j\in\cNinik \text{ and } i\ne j,\\
1-\zeta d_k(i), &\text{if } i=j,
\end{cases}$$
where 
$d_k(i)=|\cNinik|$. Additionally, let $\zeta=\tfrac{0.5}{\max_{i,k}\{d_k(i)\}}$.

The exact NE $x^*$ can be determined using the update in \eqref{eq-agent-fixed-point} with full information access.
In the partial-decision information scenario, Algorithm 1 and \cite[Alg. 1]{Bianchi2020NashES} are implemented.
The algorithms terminate when either  $\|\bz_{k+1}-\bz_k\|_\infty$ is sufficiently small or the number of iterations exceeds thresholds.

\begin{figure}[t!]
	\centering
    \includegraphics[width=0.24\textwidth]{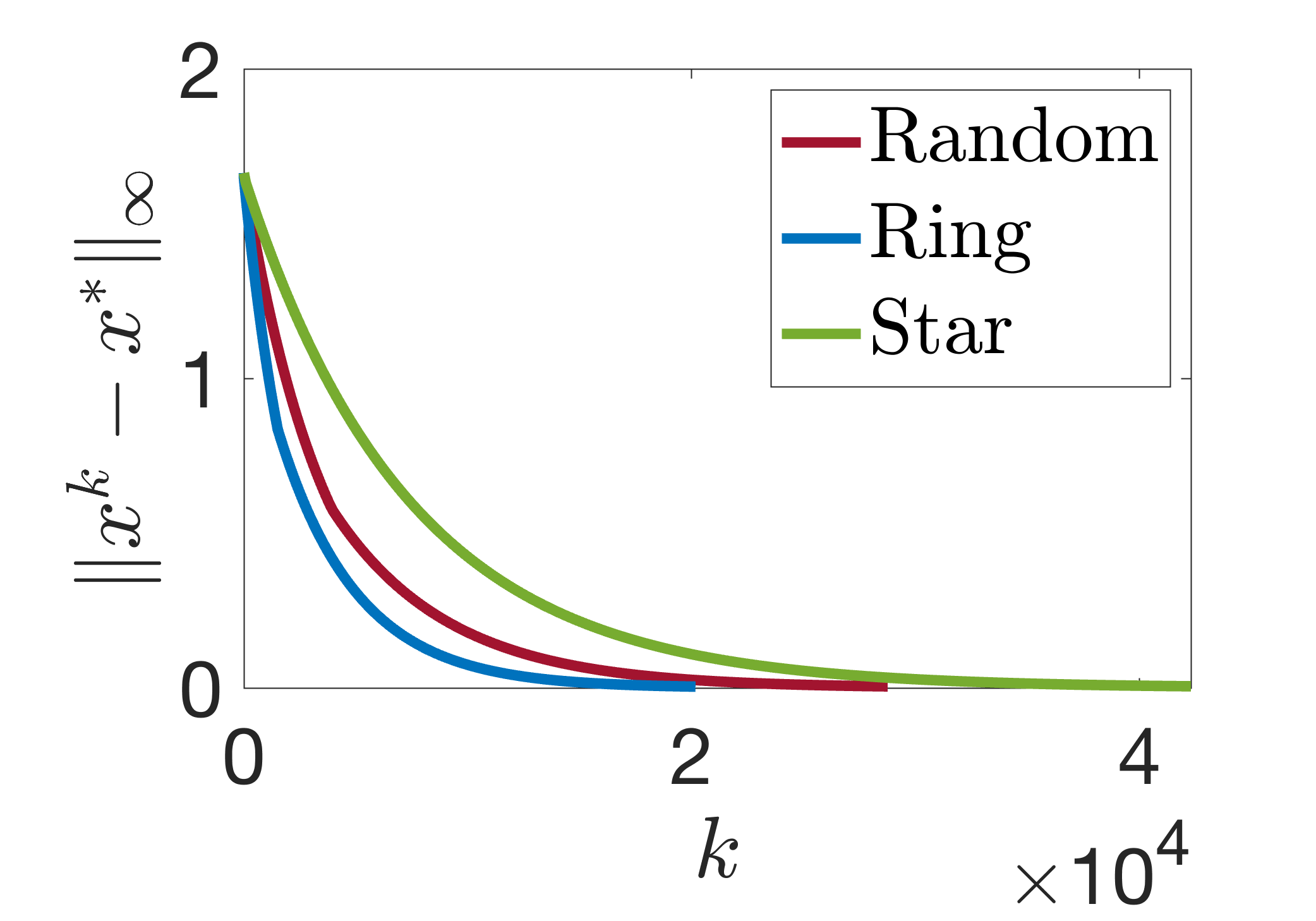} \hfill
    \includegraphics[width=0.24\textwidth]{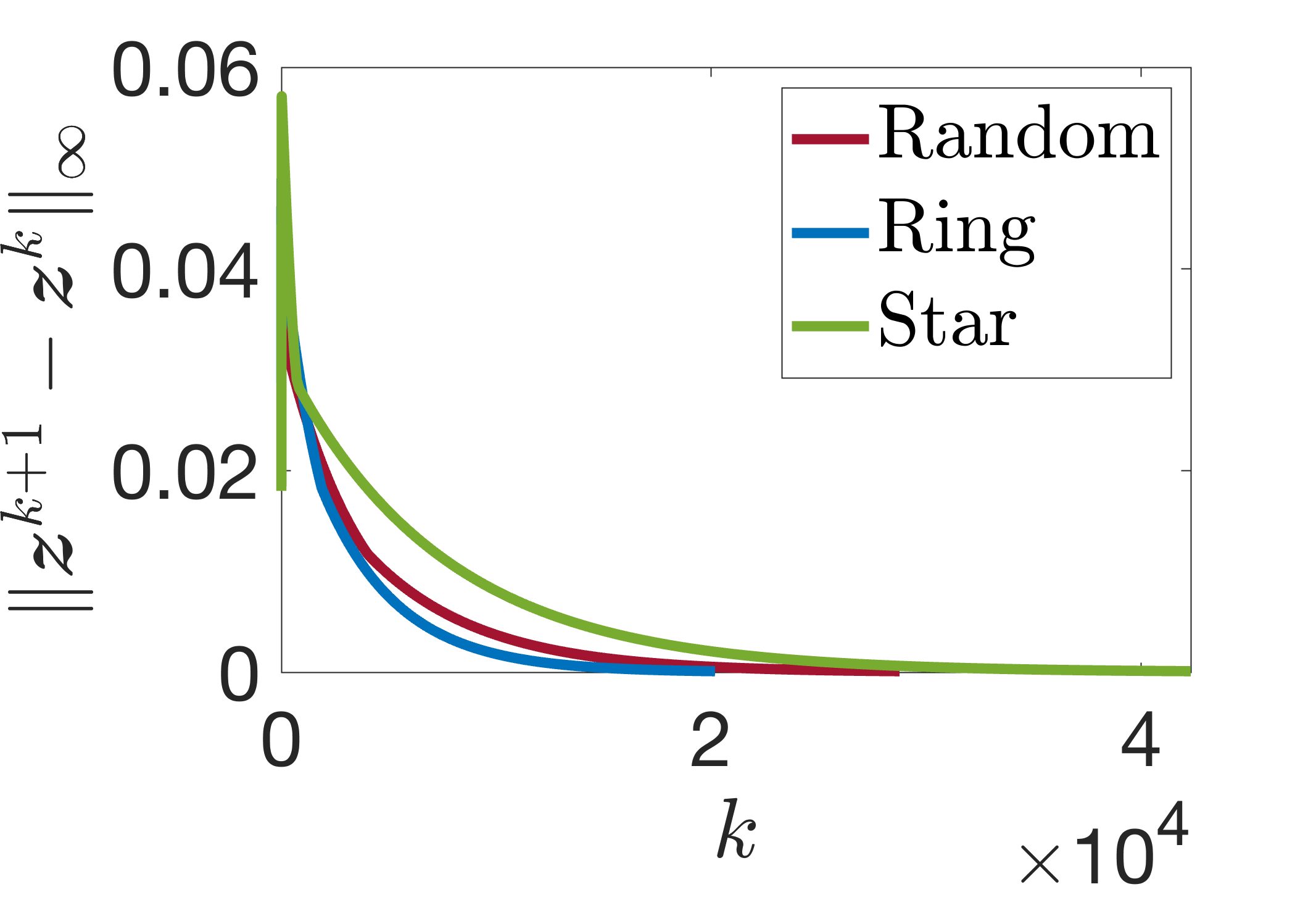}
    \vspace{-0.6cm}
	\caption{Comparison across different \textit{static} network topologies}
		\label{fig:errorsStatic}
		\vspace{-0.6cm}
\end{figure}

\begin{figure}[t!]
	\centering
	\includegraphics[width=0.24\textwidth]{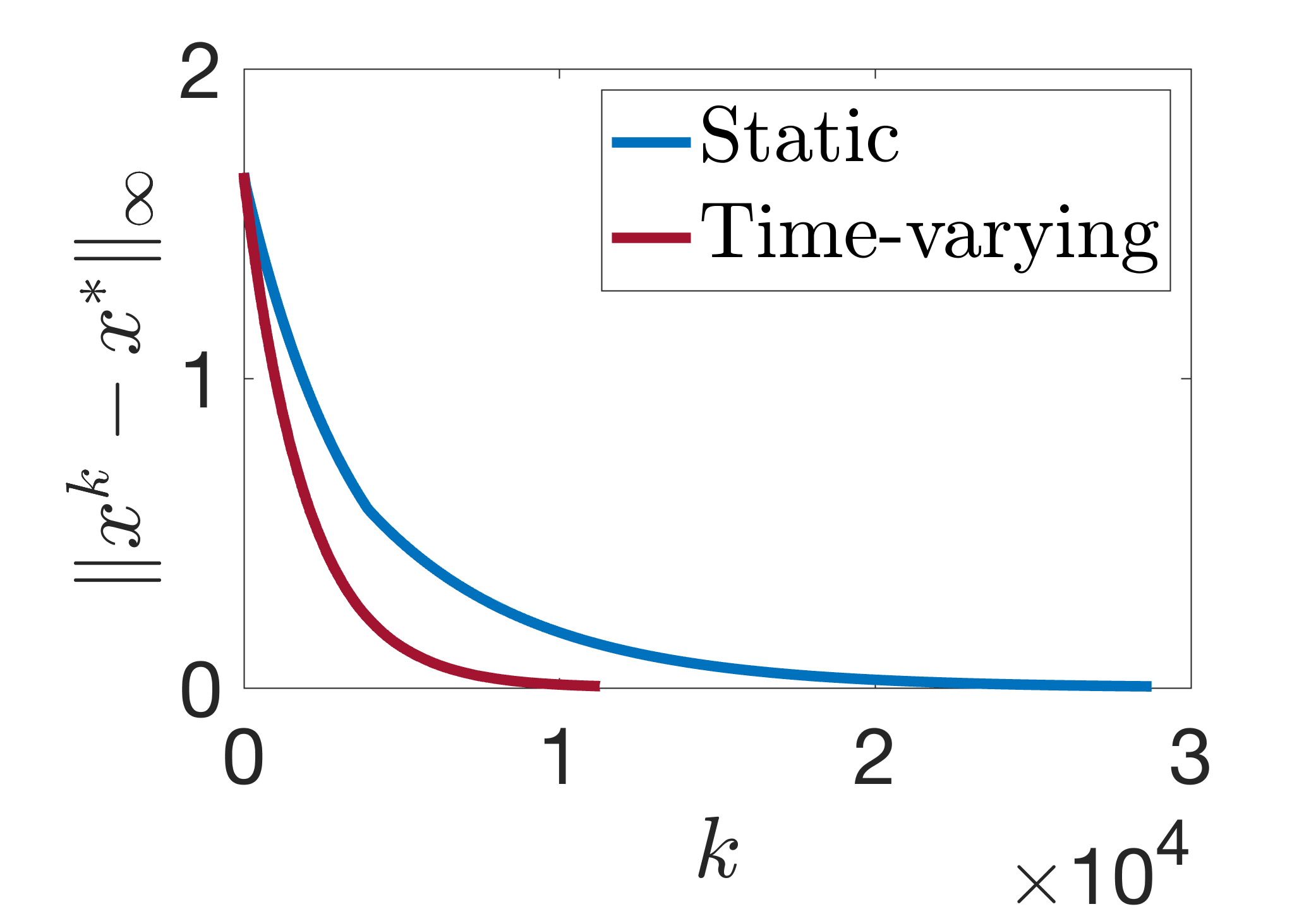}\hfill
	\includegraphics[width=0.24\textwidth]{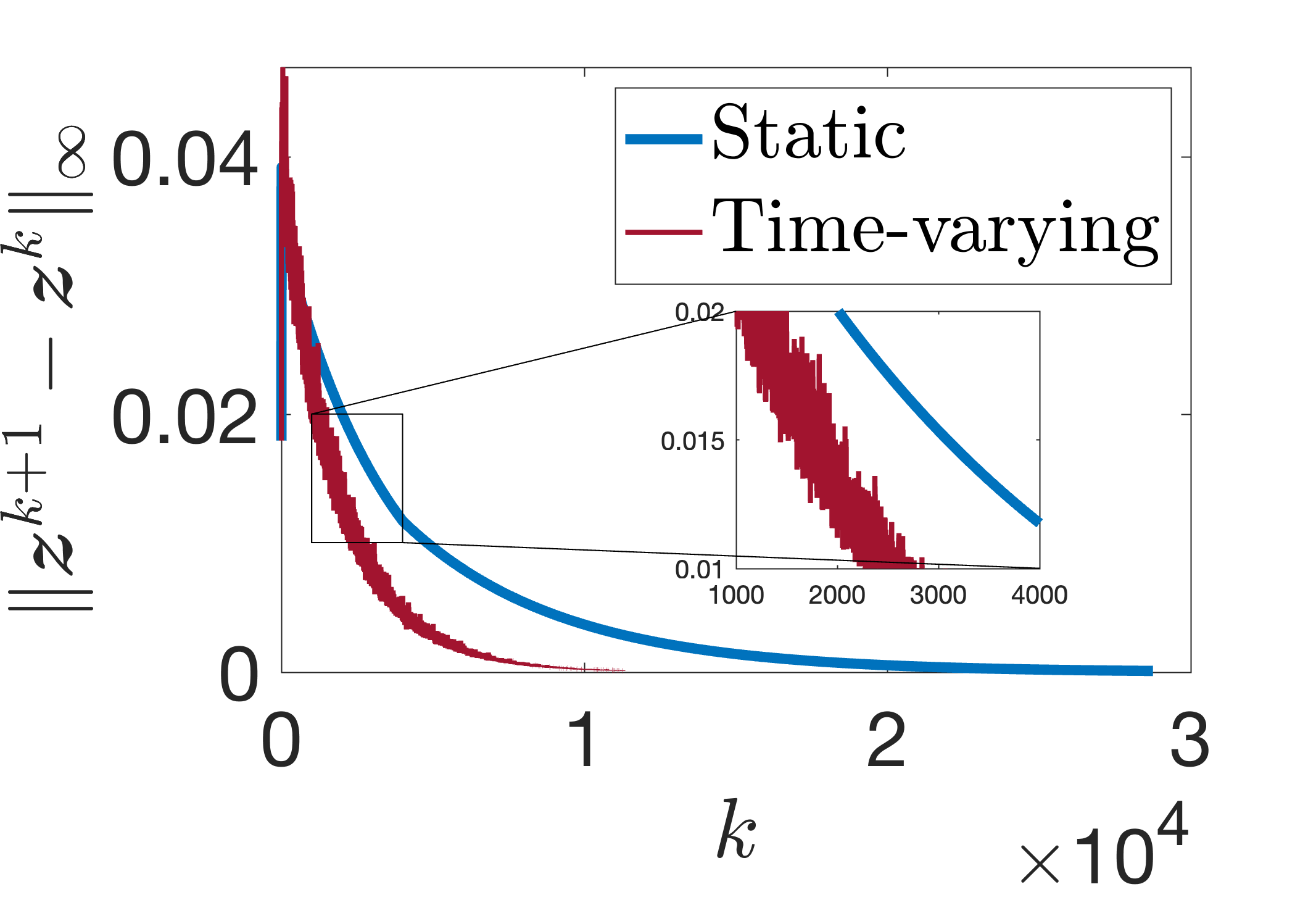}
    \vspace{-0.2cm}
	\caption{Comparing  static and time-varying directed networks.} 
	\label{fig:errorsRandom}
	\vspace{-0.4cm}
\end{figure}

\begin{figure}[t!]
	\centering
	\includegraphics[width=0.24\textwidth]{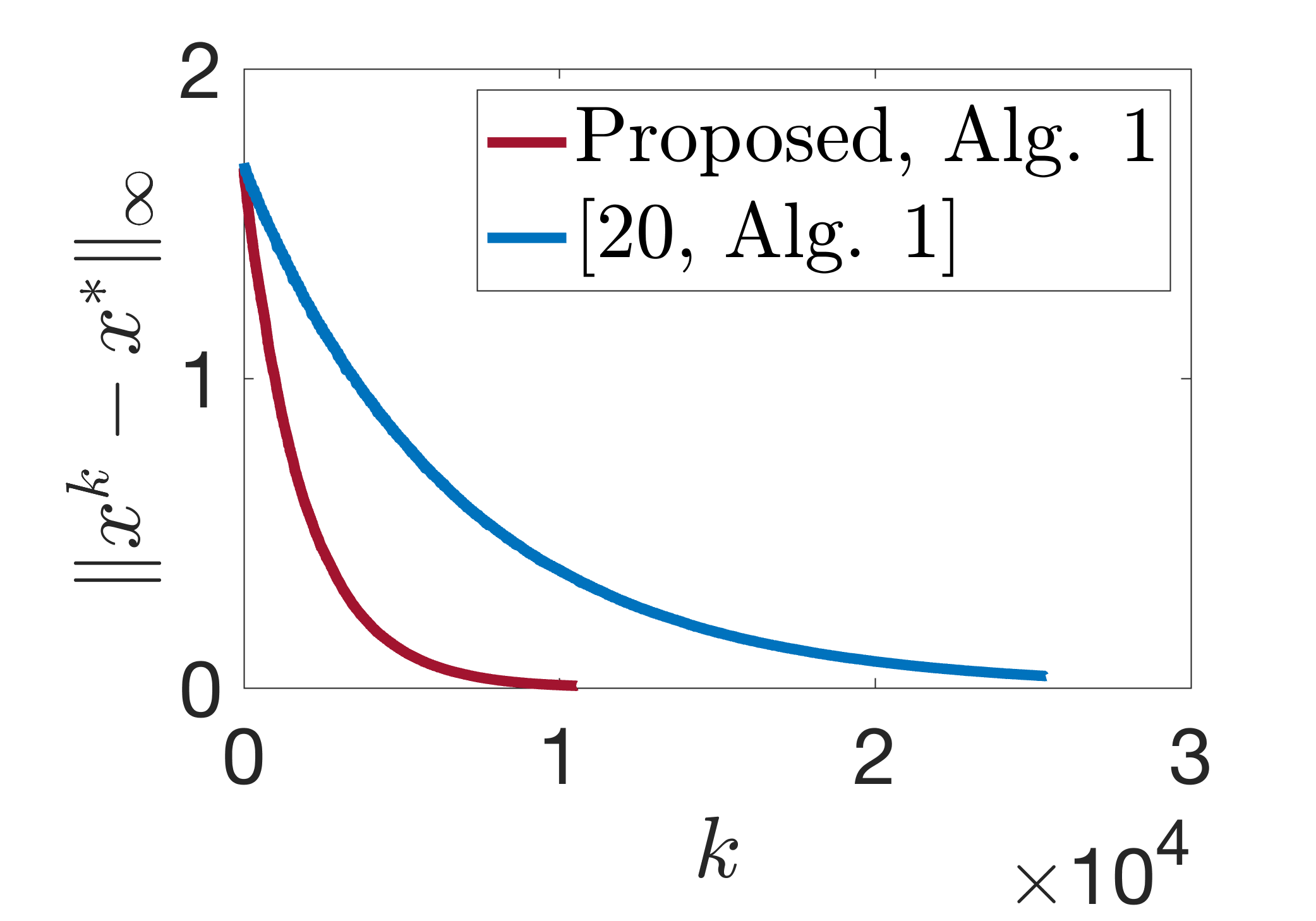} \hfill
	\includegraphics[width=0.24\textwidth]{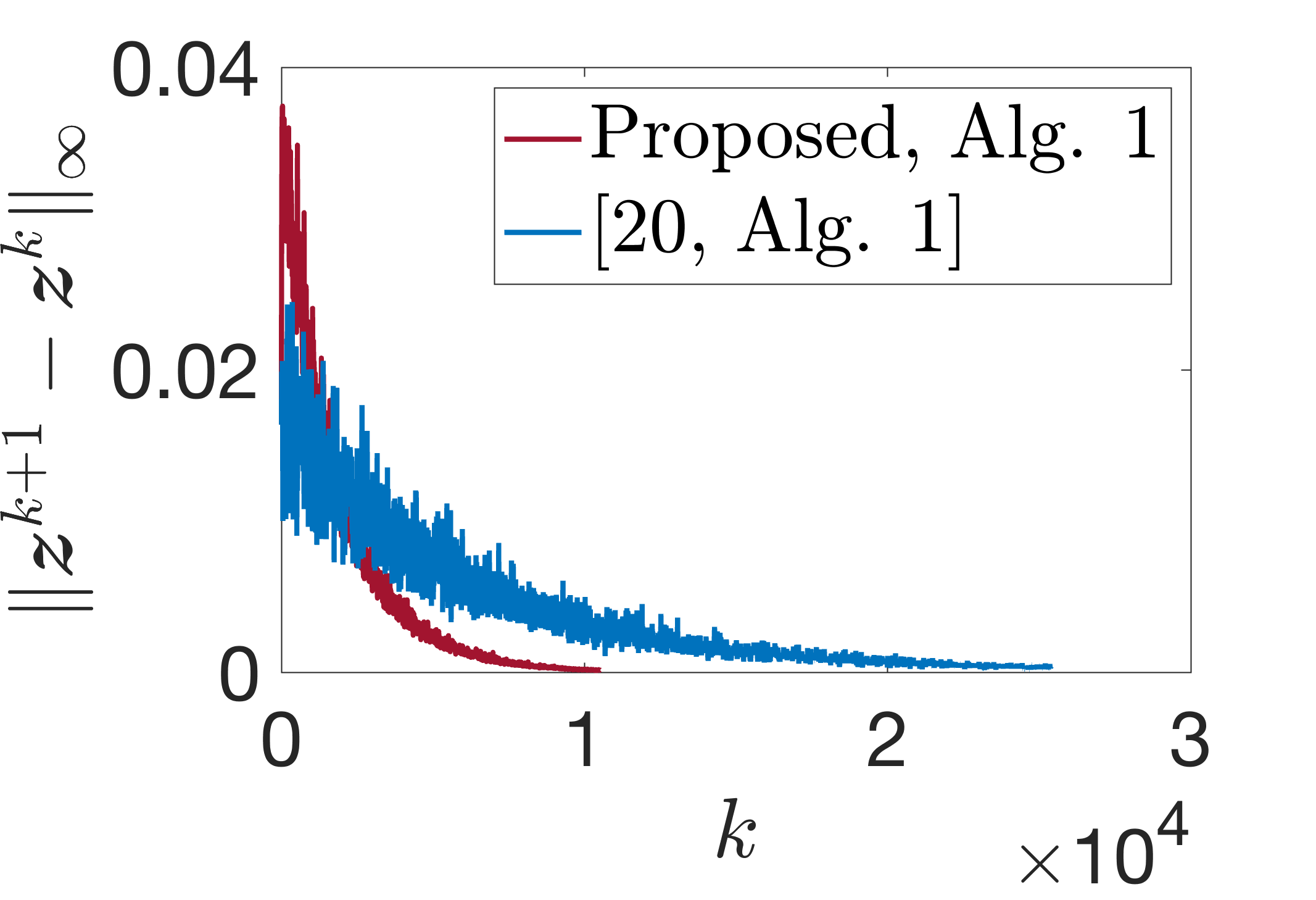}
    \vspace{-0.6cm}
	\caption{Comparing the proposed algorithm with \cite[Alg. 1]{Bianchi2020NashES}.}
	\label{fig:errors}
	\vspace{-0.6cm}
\end{figure}

Fig.~\ref{fig:errorsStatic} and Table~\ref{table:Algo1CompareGraphs} illustrate the convergence properties of Algorithm 1 over different \textit{static} communication graphs as depicted in Fig.~\ref{fig:Communication graph}. We use an identical constant stepsize of $\alpha = 0.0005$. The convergence rate is observed to be fastest for the ring connectivity and slowest for the star-shaped network. This aligns with expectations as, in the star-shaped network, all information must pass through a central firm before reaching other agents, thus increasing the number of communications required for information exchange among the firms.}

Fig.~\ref{fig:errorsRandom} compares the convergence of Algorithm 1 under static and time-varying directed communication networks. An interesting observation is the faster convergence of the method for time-varying networks. Intuitively, the variation in connections among all the firms provides more opportunities for information exchange and updates. 

For \textit{time-varying directed networks}, Fig.~\ref{fig:errors} illustrates the convergence performance of Algorithm 1 compared to \cite[Alg. 1]{Bianchi2020NashES} for a single setting. Table~\ref{table:Algo1vsBianchi1000Simulations} presents the average performance across $1000$ simulations with varying settings. A key advantage of our algorithm is that it does not require scaling the stepsizes by the PF eigenvector, a necessity for \cite{Bianchi2020NashES}. This advancement significantly reduces computational overhead, as shown in both  Fig.~\ref{fig:errors} and the computational time in Table~\ref{table:Algo1vsBianchi1000Simulations}. In fact, computing the PF eigenvector as in \cite{Bianchi2020NashES} requires certain global knowledge at every iteration, which can be impractical, particularly in distributed settings.

The error $\|\bz_{k+1}-\bz_k\|_\infty$ (right panel of Figs.~\ref{fig:errorsStatic}--\ref{fig:errors}) quantifies the variation between consecutive iterations.
In the early iterations, this error displays more pronounced fluctuations as the algorithm performs more substantial corrections. Such fluctuations are expected and indicative of the algorithm's refinement process. 
Conversely, the error $\|x^k - x^*\|_{\infty}$ (left panel) measures the deviation from the NE and demonstrates a monotonically decreasing trend without fluctuations.

\section{Conclusion}
\label{sec:conc}
We proposed a distributed algorithm integrating the projected gradient method with decision estimation alignment through consensus for NE seeking under a partial-decision information scenario. In this approach, each agent performs a gradient step to minimize its cost function utilizing the information exchanged locally with neighbors over a time-varying directed communication network. We rigorously established the geometric convergence of the algorithm towards the NE by introducing two novel auxiliary results, followed by the derivation of a novel consensus contraction relation for mixing terms related to row-stochastic mixing matrices.
We numerically demonstrated the convergence property of the algorithm through a Nash-Cournot game.
Future research directions include exploring convergence under relaxed or alternative assumptions and extending the method to more complex generalized game models incorporating local and coupling constraints among agents' decisions.


\begin{table}[t!]
\centering
\begin{tabular}{|l|c|c|c|}
\hline
\multicolumn{1}{|c|}{Graph Type}                              &\hspace{-0.15cm} \# Iterations\hspace{-0.15cm} & Run Time (s) & \hspace{-0.15cm}$\|x^k\!-x^*\|_\infty\!$\hspace{-0.15cm} \\ \hline
\hspace{-0.15cm}Static Random\hspace{-0.15cm}                                                 & $28754$          & $13.291$                                                      & $9.4\times 10^{-7}$    \\ \hline
\hspace{-0.15cm}Static Ring                                                   & $20173$ & $6.251$                                                       & $1.2\times 10^{-6}$           \\ \hline
\hspace{-0.15cm}Static Star                                                   & $42298$          & $14.620$                                                       & $7.1\times 10^{-7}$     \\ \hline
\hspace{-0.15cm}TV Random & $11266$           & $5.215$                                                       & $1.0\times 10^{-5}$    \\ \hline
\end{tabular}
\caption{Average performance of Algorithm 1 compared across different communication topologies.} 
\label{table:Algo1CompareGraphs}
\vspace{-0.3cm}
\end{table}

\begin{table}[t!]\centering
\begin{tabular}{|c|c|c|c|c|}
\hline
\multirow{2}{*}{$\a$} & \multicolumn{2}{c|}{\# Iterations}                                                                            & \multicolumn{2}{c|}{Run Time (s)}   \\ \cline{2-5} 
&\!\!Proposed, Alg. 1\!\!    & \!\!\!\!\cite[Alg. 1]{Bianchi2020NashES}\!\!  &  \!\!Proposed, Alg. 1\!\!  & 	\!\!\!\! \cite[Alg. 1]{Bianchi2020NashES}\!\!  \\ \hline
\!\!\! $0.0005$\!\!\!  & \!$10403.93 $\! & \!\! \!$25358.28$\! & \!$4.6161 $\! & \!\! \!$29.5852 $\!       \\ \hline
\!\!\! $0.0001$\!\!\!  & \!$18045.61 $\! & \!\! \!$42082.22 $\! & \!$8.1566 $\! & \!\! \!$49.8984$\!     \\ \hline
\end{tabular}
\caption{Comparison of average performance between the proposed algorithm and \cite[Alg. 1]{Bianchi2020NashES}.}
\label{table:Algo1vsBianchi1000Simulations} 
\vspace{-0.6cm}
\end{table}


\bibliographystyle{IEEEtran}
\bibliography{references}

\newpage
\appendix
\subsection{Proof of Lemma~\ref{lem-normlincomb}}\label{app:lem-normlincomb}
\begin{proof}
(a)~For $\|\sum_{i=1}^m \g_i u_i\|^2$, we have:
	\begin{align*}
	\left\|\sum_{i=1}^m \g_i u_i\right\|^2=\sum_{i=1}^m \g_i^2 \|u_i\|^2 
	+ \sum_{i=1}^m \sum_{j=1, j\ne i}^m \g_i \g_j \la u_i,u_j\ra.
	\end{align*}
	Using the identity $2\la v,w\ra=\|v\|^2+\|w\|^2-\|v-w\|^2$, yields:
	\begin{align}\label{eq-a1}
	\left\|\sum_{i=1}^m \g_i u_i\right\|^2\!\!
	&=\sum_{i=1}^m \g_i^2 \|u_i\|^2\\
    &+ \frac{1}{2}
	\sum_{i=1}^m \!\sum_{j=1, j\ne i}^m\!\!\! \g_i \g_j \!\left(\|u_i\|^2\!+\!\|u_j\|^2\!-\!\|u_i\!-\!u_j\|^2\right)\!.\nonumber
	\end{align} 
 Note that
	\begin{align*}
	&\sum_{i=1}^m \sum_{j=1, j\ne i}^m \g_i \g_j \left(\|u_i\|^2+\|u_j\|^2\right)
 	\\
  =&\sum_{i=1}^m \Bigg(\sum_{j=1, j\ne i}^m \g_j\Bigg) \!\g_i \|u_i\|^2
 	+\!\sum_{j=1}^m \!\Bigg(\sum_{i=1, i\ne j}^n \!\g_i \!\Bigg)\g_j \|u_j\|^2\!.
 	\end{align*}
 	We further obtain,
 	\[\sum_{i=1}^m  \sum_{j=1, j\ne i}^m \!\! \g_i \g_j \left(\|u_i\|^2\!+\|u_j\|^2\right)
 	\!=\!2\! \sum_{i=1}^m \!\Bigg(\sum_{j=1, j\ne i}^m \!\g_j\!\!\Bigg)\g_i \|u_i\|^2\!.\]
    Substituting the preceding equality into relation~\eqref{eq-a1}, we find:
 	\begin{align*}
 	\left\|\sum_{i=1}^m \g_i u_i\right\|^2
 	&=\sum_{i=1}^m \g_i^2 \|u_i\|^2 
 	+\sum_{i=1}^m \Bigg(\sum_{j=1, j\ne i}^m \g_j\Bigg) \g_i \|u_i\|^2\\
    &-\frac{1}{2}\sum_{i=1}^m \sum_{j=1, j\ne i}^m \g_i \g_j \|u_i-u_j\|^2.
 	\end{align*}
	The relation in part (a) follows by combining the first two terms and noting that $\g_i\g_j\|u_i-u_j\|^2=0$ for $j=i$.
	
	\noindent
	(b)~Suppose now that $\sum_{i=1}^m \g_i=1$. For any $u\in\re^n$,
 	we have:
	 \[\sum_{i=1}^m \g_i u_i - u
 	= \sum_{i=1}^m \g_i u_i - \left(\sum_{i=1}^m \g_i\right) u
	= \sum_{i=1}^m \g_i (u_i-u).\]
    Using the relation from part (a) with $u_i$ replaced by $u_i - u$ and $\sum_{i=1}^m \g_i = 1$, we derive the desired result in part (b).
 \end{proof}

\vspace{-0.2cm}
\subsection{Proof of Lemma~\ref{lem-xdiverse}}\label{app-lem-xdiverse}
\begin{proof}
	Let $\mathcal{P}^*\in\mathcal{S}(\bbG)$ be a shortest path covering of the graph $\bbG$. Specifically, $\mathcal{P}^*=\{\bp^*_{j\ell}\mid j,\ell\in[m], j\ne\ell\},$ where $\bp^*_{j\ell}$ is a shortest path from node $j$ to node $\ell$. 
	Let $\E^*$ be the collection of all directed links that are traversed by any path in  $\mathcal{P}^*$, i.e.,
	$\E^*\!=\!\{(i,q)\in\E\mid (i,q)\in\bp_{j\ell}^*\ \hbox{ for some }  \bp_{j\ell}^*\in\mathcal{P}^*\}.$
 
	By Definition~\ref{def-edgeut} of the maximal edge-utility $\mathsf{K}(\bbG)$, we have
	$\mathsf{K}(\bbG)\ge \mathsf{K}(\mathcal{P}^*),$
	where $\mathsf{K}(\mathcal{P}^*)$ is the maximal edge-utility with respect to the shortest path 
	covering $\mathcal{P}^*$, i.e.,
	\[\mathsf{K}(\mathcal{P}^*)
	=\max_{(i,q)\in\E} \sum_{\bp\in\mathcal{P}^*}\chi_{\{(i,q)\in \bp\}}.\]
	Note that $\sum_{\bp\in\mathcal{P}^*}\chi_{\{(i,q)\in \bp\}}=0$ 
	when a link $(i,q)$ is not used by any of the paths in $\mathcal{P}^*$. Thus, the value
	$\mathsf{K}(\mathcal{P}^*)$ is equivalently given by:	
	$ \mathsf{K}(\mathcal{P}^*)
	=\max_{(i,q)\in\E^*}\sum_{\bp\in\mathcal{P}^*}\chi_{\{(i,q)\in \bp\}}$.
	
	Given that $\E^*\subseteq\E$, we have
	\begin{align}\label{eq-pathest1}
	&\mathsf{K}(\mathcal{P}^*)\sum_{(j,\ell)\in\E} \|z_j - z_\ell \|^2 \ge 
	\mathsf{K}(\mathcal{P}^*)\sum_{(i,q)\in \E^*} \|z_i - z_q \|^2 \\
    =&\!\!\!\sum_{(i,q)\in \E^*}\!\!\! \mathsf{K}(\mathcal{P}^*)\|z_i \!-\! z_q \|^2\ge \!\!\!\!\! \sum_{(i,q)\in\E^*} \!\!\! \bigg(\sum_{\bp\in\mathcal{P}^*}\!\!\chi_{\{(i,q)\in \bp\}}\!\!\bigg)
	\|z_i \!-\! z_q \|^2\!, \nonumber
	\end{align}
	since $\mathsf{K}(\mathcal{P}^*)$ is the maximal edge-utility with respect to paths in $\mathcal{P}^*$, it follows that
	$\mathsf{K}(\mathcal{P}^*)\ge \sum_{\bp\in\mathcal{P}^*}\chi_{\{(i,q)\in \bp\}}$ for any link $(i,q)$ used by a path in $\mathcal{P}^*$.
    The last sum in \eqref{eq-pathest1} 
	is over all links in $\E^*$, weighted by the multiplicity with which each $(i,q)$ appears in 
    $\mathcal{P}^*$. Thus, it can be written in terms of the paths in 
	$\mathcal{P}^*$ connecting distinct nodes $j,\ell\in[m]$,
	as follows:
	\begin{align}\label{eq-11}
	&\sum_{(i,q)\in\E^*}
	\bigg(\sum_{\bp\in\mathcal{P}^*}\chi_{\{(i,q)\in \bp\}}\bigg)
	\|z_i - z_q \|^2
	\\
	= &\sum_{j=1}^m \sum_{\ell=j+1}^m \sum_{(i,q)\in \bp_{j\ell}^*} \!\!\!\|z_i - z_q \|^2
	+\sum_{j=1}^m \sum_{\ell=j+1}^m \sum_{(i,q)\in \bp_{\ell j}^*} \!\!\!\|z_i - z_q \|^2.\nonumber
    \end{align}
	Using the convexity of the squared norm, we have:
	\begin{align*}
	\frac{1}{|\bp^*_{j\ell}|}  \!\sum_{(i,q)\in \bp_{j\ell}^*} \!\!\!\!\|z_i \!-\! z_q \|^2\!\ge\!
	 \bigg\|\frac{1}{|\bp^*_{j\ell}|}\!\sum_{(i,q)\in \bp_{j\ell}^*} \!\!\!\!(z_i \!-\! z_q)\bigg\|^2\!\!=\!\frac{\|z_j\!-\!z_\ell\|^2}{|\bp^*_{j\ell}|^2},
	 \end{align*}
	 where $|\bp^*_{j\ell}|$ denotes the length of the path $\bp^*_{j\ell}$.
	 Hence,
	 \[\sum_{(i,q)\in \bp_{j\ell}^*} \|z_i - z_q \|^2\ge \frac{1}{|\bp^*_{j\ell}|}\|z_j-z_\ell\|^2,\]
	 which in view of~\eqref{eq-11} implies that
	 \begin{align*}
	 &\sum_{(i,q)\in\E^*}
	\!\!\!\bigg(\sum_{\bp\in\mathcal{P}^*}\!\!\chi_{\{(i,q)\in \bp\}}\! \bigg) \|z_i \!-\! z_q \|^2\\
	\ge&\sum_{j=1}^m\sum_{\ell=j+1}^m \!\frac{1}{|\bp^*_{j\ell}|}\,\! \|z_j \!-\! z_\ell\|^2\!+\!\sum_{j=1}^m\sum_{\ell=j+1}^m \!\frac{1}{|\bp^*_{\ell j}|}\,\! \|z_j \!-\! z_\ell\|^2\!.
	\end{align*}
	Hence, by Definition~\ref{def-diam} of $\mathsf{D}(\bbG)$, we have $|\bp^*_{\ell j}|\le \mathsf{D}(\bbG)$ for any $j\ne \ell$, implying that
	\begin{align*}
	\sum_{(i,q)\in\E^*}\!\!\!\bigg(\!\!\!\!\!\sum_{~~\bp\in\mathcal{P}^*}\!\!\!\chi_{\{(i,q)\in \bp\}}\!\!\bigg)\!
	\|z_i \!-\! z_q \|^2
	\ge \frac{2}{\mathsf{D}(\bbG)}\!\sum_{j=1}^m\sum_{\ell=j+1}^m  \!\!\|z_j \!-\! z_\ell\|^2\!.
	\end{align*}
	Using the relation in~\eqref{eq-pathest1} and the above relation,
	we obtain
	\[\sum_{(j,\ell)\in\E} \|z_j - z_\ell \|^2
	\ge \frac{2}{\mathsf{D}(\bbG)\mathsf{K}(\mathcal{P}^*)}\sum_{j=1}^m\sum_{\ell=j+1}^m\|z_j - z_\ell\|^2.\]
	The desired relation follows from $\mathsf{K}(\bbG)\ge \mathsf{K}(\mathcal{P}^*).$
	\end{proof}

\vspace{-0.2cm}
\subsection{Proof of Lemma~\ref{lemma-TimeVaryPiVectors}}\label{app-lemma-TimeVaryPiVectors}
\begin{proof} 
By Assumption~\ref{asum-graphs}, the sequence $\{W_k\}$ of row-stochastic matrices is ergodic. Thus, 
Theorem 4.20 of~\cite{Seneta} on backwards products implies that there exists a unique sequence of absolute probability vectors $\{\pi_k\}$, i.e., a sequence $\{\pi_k\}$ of stochastic vectors such that $\pi_{k+1}'W_k=\pi_k'$ for all $k\ge 0.$ This shows the result in part (a).
The statement in part (b) follows from Lemma 4 and Remark 2 of~\cite{Nedic2015}. Specifically, in the proof of \cite[Lemma 4]{Nedic2015}, using the lower bound $[W_k]_{ii}\ge w$ and $B=1$, we obtain $\delta'\ge w^m$. Noting that the vector sequence $\{\phi(k)\}$ in \cite{Nedic2015} coincides with the sequence $\{\pi_k\}$ in this work (see the proof of \cite[Lemma 1]{Angelia2017}), \cite[Remark 2]{Nedic2015} then gives $[\pi_k]_i\ge\frac{w^m}{m}$ for all $i\in[m]$ and all $k\ge0$.
\end{proof}

\subsection{Proof of Lemma~\ref{lemma-LipschitzbF}} \label{app-lemma-LipschitzbF}
\begin{proof}
We apply Lemma~\ref{lemma:LipschitzNablaJi}, as follows
\begin{align*} 
&\|\bF_{\bda}{(\bx)}-\bF_{\bda}{(\by)}\|_\pi^2 
= \sum_{i=1}^m\pi_i \a_i^2 \left\lVert\nabla_i J_i(x_{i:})-\nabla_i J_i(y_{i:})\right\rVert^2\\
\le& \!\sum_{i=1}^m\!\pi_i \a_i^2\!\!\left(L_{-i}^2 \!+\! L_i^2\right)  \!\! \left\lVert x_{i:} \!-\! y_{i:}\right\rVert^2
\!\!\le\! \max\limits_{i\in[m]}\!\!\left\{\! \a_i^2\!\left(L_{-i}^2 \!+\! L_i^2\right) \!\right\}\!\! \|\bx \!-\! \by\|_\pi^2
\end{align*}
which completes the proof.
\end{proof}

\vspace{-0.2cm}
\subsection{Proof of Corollary~\ref{corollary-lemma6time}}\label{app-lemma6time}
\revtwo{
\begin{proof}
Since all components in $\eta_k$ are positive, $\eta_k > 0$ holds trivially. To establish that $\eta_k < 1$, we note:
\begin{list}{\labelitemi}{\leftmargin=5pt \itemsep=0pt \parsep=0pt}
    \item $ 0 < \tfrac{w^m}{m} \leq \min(\pi_{k+1}) \leq \tfrac{1}{m}\leq \max(\pi_k) < 1.$
    \item As $\bbG_k$ is strongly connected, we have: $1 \leq \mathsf{D}(\bbG_k) \leq m - 1$. 
    \item Since $W_k$ is row-stochastic and the maximum out-degree in the graph, $\max_{i \in [m]} |\cNoutik|$, is at least the average out-degree per node, $s_k = \left\lceil \frac{|\E_k|}{m} \right\rceil$, we have: 
    $w \leq \tfrac{1}{\max_{i\in [m]} |\cNoutik|} \leq \tfrac{1}{s_k}$.
    \item  $\mathsf{K}(\bbG_k)$ is at least the average number of shortest paths passing through an edge. With $m(m - 1)$ shortest paths and $|\E_k| - m$ edges (excluding self-loops): $\mathsf{K}(\bbG_k) \geq \frac{m(m - 1)}{|\E_k| - m}$.
\end{list}
Using these bounds, we have:
    \begin{align*}
    \eta_k & \leq \tfrac{\left( \tfrac{1}{m} \right) \left( \tfrac{1}{s_k} \right)^2}{\left( \tfrac{1}{m} \right)^2 \cdot 1 \cdot \left( \tfrac{m(m - 1)}{|\E_k| - m} \right)}  = \dfrac{|\E_k| - m}{(m - 1) s_k^2} < \frac{m s_k - 1}{(m - 1) s_k^2} < 1,
    \end{align*}
where $s_k = \left\lceil \frac{|\E_k|}{m} \right\rceil$ implies $ |\E_k| - m < m s_k - 1 $, and
\[
(m - 1) s_k^2 - (m s_k - 1) = (s_k - 1)[s_k (m - 1) - 1] > 0,
\]
for $m \geq 2$ and $s_k \geq 2$ (since $\bbG_k$ is strongly connected and has self-loops at every node).
\end{proof}
}

\vspace{-0.2cm}
\subsection{Proof of Lemma~\ref{lem-basic}}\label{app-lem-basic}
\begin{proof}
According to Algorithm 1 and notations in~\eqref{eq-notat}:
\begin{align}\label{eq:ProjNorm1}
&\|\bz^{k+1}-\bx^*\|_{\pi_{k+1}}^2 \cr
=&\sum_{i=1}^m[\pi_{k+1}]_i\left(\|x_i^{k+1}-x_i^*\|^2 + \|z_{i,-i}^{k+1}-x_{-i}^*\|^2\right).
\end{align}
From relation \eqref{eq-agent-fixed-point} and projection non-expansiveness, follows:
\begin{align}\label{eq:ProjNorm2}
\!\!&\|x_i^{k+1}-x_i^*\|^2\nonumber \\
\!\!=&\|\Pi_{X_i}\!\!\left[v_{ii}^{k+1} \!\!-\! \a_i^{k+1}\nabla_i J_i(v_{i}^{k+1})\!\right] \!\!-\!\Pi_{X_i}\![x_i^* \!\!-\! \a_i^{k+1} \nabla_{i} J_i(x^*)]\|^2\nonumber \\
\!\!\le & \| v_{ii}^{k+1} \!-\a_i^{k+1}\nabla_i J_i(v_{i}^{k+1})-
(x_i^*\!-\a_i^{k+1} \nabla_i J_i(x^*))\|^2\!.
\end{align}
Combining \eqref{eq:ProjNorm1}, \eqref{eq:ProjNorm2}, and the updates for 
$v_{i}^{k+1}$ yields
\begin{align}
    &\|\bz^{k+1}-\bx^*\|_{\pi_{k+1}}^2 \nonumber\\
    \le& \|(W_k\bz^k-\bx^*)-(\bF_{\bda^{k+1}}{(W_k\bz^k)}-\bF_{\bda^{k+1}}{(\bx^*)})\|_{\pi_{k+1}}^2\nonumber\\
    =& \|W_k\bz^k-\bx^*\|_{\pi_{k+1}}^2+\|\bF_{\bda^{k+1}}{(W_k\bz^k)}-\bF_{\bda^{k+1}}{(\bx^*)}\|_{\pi_{k+1}}^2\nonumber\\
    &-2\la W_k\bz^k -\bx^*,\bF_{\bda^{k+1}}{(W_k\bz^k)} -\bF_{\bda^{k+1}}{(\bx^*)} \ra_{\pi_{k+1}}\nonumber\\
    \le &(1+\bL_{\a} ^2) \|W_k\bz^k-\bx^*\|_{\pi_{k+1}}^2\cr
    & -2\la W_k\bz^k-\bx^*,\bF_{\bda^{k+1}}{(W_k\bz^k)}-\bF_{\bda^{k+1}}{(\bx^*)} \ra_{\pi_{k+1}}~~~ \label{eq:piNormF}
\end{align}
where the last inequality follows from Lemma~\ref{lemma-LipschitzbF}:
\begin{align*} 
\|\bF_{\bda^{k+1}}{(W_k\bz^k)}-\bF_{\bda^{k+1}}{(\bx^*)}\|_{\pi_{k+1}}^2 \le \bL_{\a} ^2\|W_k\bz^k-\bx^*\|_{\pi_{k+1}}^2.
\end{align*}
To estimate the inner product in \eqref{eq:piNormF}, we write 
\begin{align} 
    &\la W_k\bz^k-\bx^*,\bF_{\bda^{k+1}}{(W_k\bz^k)}-\bF_{\bda^{k+1}}{(\bx^*)} \ra_{\pi_{k+1}}\nonumber\\
    =& \la W_k\bz^k-\bx^*,\bF_{\bda^{k+1}}{(W_k\bz^k)}-\bF_{\bda^{k+1}}{(\hat{\bz}_{\pi_{k}})} \ra_{\pi_{k+1}}\nonumber \\
    &+\la W_k\bz^k-\hat{\bz}_{\pi_{k}},\bF_{\bda^{k+1}}{(\hat{\bz}_{\pi_{k}})}-\bF_{\bda^{k+1}}{(\bx^*)} \ra_{\pi_{k+1}} \nonumber \\
    &+\la \hat{\bz}_{\pi_{k}}-\bx^*,\bF_{\bda^{k+1}}{(\hat{\bz}_{\pi_{k}})}-\bF_{\bda^{k+1}}{(\bx^*)} \ra_{\pi_{k+1}}. \label{eq:innerProd}
\end{align}
Applying the Cauchy–Schwarz inequality and Lemma~\ref{lemma-LipschitzbF} yields
\begin{align}\label{eq-est1}
    &|\la W_k\bz^k \!-\!\bx^*,\bF_{\bda^{k+1}}{(W_k\bz^k)}\!-\!\bF_{\bda^{k+1}}{(\hat{\bz}_{\pi_{k}})} \ra_{\pi_{k+1}}| \nonumber\\
    \!\le& \|W_k\bz^k \!-\bx^*\|_{\pi_{k+1}}\|\bF_{\bda^{k+1}}{(W_k\bz^k)} \!-\!\bF_{\bda^{k+1}}{(\hat{\bz}_{\pi_{k}})}\|_{\pi_{k+1}} \nonumber\\
    \le& \bL_{\a} \|W_k\bz^k-\bx^*\|_{\pi_{k+1}}\|W_k\bz^k-\hat{\bz}_{\pi_{k}}\|_{\pi_{k+1}}.
\end{align}
Similarly,
\begin{align}\label{eq-est2}
    &|\la W_k\bz^k-\hat{\bz}_{\pi_{k}},\bF_{\bda^{k+1}}{(\hat{\bz}_{\pi_{k}})}-\bF_{\bda^{k+1}}{(\bx^*)} \ra_{\pi_{k+1}}| \nonumber\\
    \le & \bL_{\a} \|W_k\bz^k-\hat{\bz}_{\pi_{k}}\|_{\pi_{k+1}}\|\hat{\bz}_{\pi_{k}}-\bx^*\|_{\pi_{k+1}}.
\end{align}

To estimate the last inner product in \eqref{eq:innerProd}, we write
\begin{align} \label{eq-innerproductdereive1} 
    &\la \hat{\bz}_{\pi_{k}}-\bx^*, \bF_{\bda^{k+1}}{(\hat{\bz}_{\pi_{k}})}-\bF_{\bda^{k+1}}{(\bx^*)} \ra_{\pi_{k+1}}\nonumber\\
    =& \sum_{i=1}^m [\pi_{k+1}]_i \a_i^{k+1} \la[\hat{z}_{\pi_{k}}]_i-x_i^*, \nabla_i J_i(\hat{z}_{\pi_{k}})- \nabla_i J_i(x^*)\ra\nonumber\\
    =& \!\sum_{i=1}^m ([\pi_{k+1}]_i\a_i^{k+1} \!\!-\!\tilde{\a}_{\pi_{k+1}}\!)\la[\hat{z}_{\pi_{k}}]_i\!-\!x_i^*, \!\nabla_i J_i(\hat{z}_{\pi_{k}}\!)\!-\! \nabla_i J_i(x^*)\ra \nonumber\\
    &+ \tilde{\a}_{\pi_{k+1}}\sum_{i=1}^m \la[\hat{z}_{\pi_{k}}]_i-x_i^*, \nabla_i J_i(\hat{z}_{\pi_{k}})- \nabla_i J_i(x^*)\ra.
\end{align}

\revtwo{For the first term in \eqref{eq-innerproductdereive1}, we have
\begin{align}\label{eq-innerproductdereive2}
&\sum_{i=1}^m ([\pi_{k+1}]_i\a_i^{k+1} \!\!-\!\tilde{\a}_{\pi_{k+1}}\!)\la[\hat{z}_{\pi_{k}}]_i\!-\!x_i^*, \!\nabla_i J_i(\hat{z}_{\pi_{k}}\!)\!-\! \nabla_i J_i(x^*)\ra\nonumber\\
\overset{\!\!\!\!\!\tiny{\text{(a)}}\!\!\!\!\!}{\ge}& \!-\!\!\!\sum_{i=1}^m \!|[\pi_{k+\!1}]_i\a_i^{k+\!1} \!\!-\!\tilde{\a}_{\pi_{k+\!1}}\!| \|[\hat{z}_{\pi_{k}}]_i\!\!-\!x_i^*\| \| \nabla_i J_i(\!\hat{z}_{\pi_{k}}\!)\!-\!\! \nabla_i J_i(\!x^*\!)\|\nonumber\\
\overset{\!\!\!\!\!\tiny{\text{(b)}}\!\!\!\!\!}{\ge}& -L\| \hat{z}_{\pi_{k}}- x^*\|\sum_{i=1}^m |[\pi_{k+1}]_i\a_i -\tilde{\a}_{\pi_{k+1}}| \|[\hat{z}_{\pi_{k}}]_i-x_i^*\| \nonumber\\
\overset{\!\!\!\!\!\tiny{\text{(c)}}\!\!\!\!\!}{\ge}& -L\| \hat{z}_{\pi_{k}}- x^*\|^2 \left\| \Diag(\pi_{k+1}) \bda^{k+1} - \tilde{\bda}_{\pi_{k+1}} \right\|\nonumber\\
\overset{\!\!\!\!\!\tiny{\text{(d)}}\!\!\!\!\!}{\ge}& - L\epsilon_{\a}^{k+1}\tilde{\a}_{\pi_{k+1}} \| \hat{z}_{\pi_{k}}- x^*\|^2,
\end{align}
where (a) applies the Cauchy-Schwarz inequality, $w \la u, v \ra \geq -|w| |\la u, v \ra| \!\geq \! -|w| \|u\| \|v\|$;
(b) uses Lemma~\ref{lemma:LipschitzNablaJi} with $L$ as defined in \eqref{eq-constants}; (c) follows from the Cauchy-Schwarz inequality $\la u, v \ra \!\leq \! \|u\| \|v\|$; and (d) uses the definition of $\epsilon_{\a}^{k+1}$.} 

For the second term in \eqref{eq-innerproductdereive1}, using Assumption~\ref{assum:map_monotone} gives
\begin{align}\label{eq-innerproductdereive3}
 &\tilde{\a}_{\pi_{k+1}}\sum_{i=1}^m \la[\hat{z}_{\pi_{k}}]_i-x_i^*, \nabla_i J_i(\hat{z}_{\pi_{k}})- \nabla_i J_i(x^*)\ra\nonumber\\
 \ge & ~\mu \tilde{\a}_{\pi_{k+1}} \| \hat{z}_{\pi_{k}}- x^*\|^2.
\end{align}

Combining the relations in \eqref{eq-innerproductdereive1}--\eqref{eq-innerproductdereive3} and noting that $\| \hat{z}_{\pi_{k}} - x^*\|^2= \|\hat{\bz}_{\pi_{k}}-\bx^*\|_{\pi_{k}}^2$, we have the following relation for the last inner product in \eqref{eq:innerProd}:
\begin{align*}
    \la \hat{\bz}_{\pi_{k}}\!\!-\!\bx^*\!, \,\bF_{\bda^{k+1}}{(\hat{\bz}_{\pi_{k}})} \!-\! \bF_{\bda^{k+1}}{(\bx^*)} \ra_{\pi_{k+1}} \!\!\ge \bdbeta_{\a} \|\hat{\bz}_{\pi_{k}}\!-\!\bx^*\|_{\pi_{k}}^2.
 \end{align*}
By substituting the estimates \eqref{eq:innerProd}--\eqref{eq-innerproductdereive3} and the preceding relation back into \eqref{eq:piNormF}, we derive the desired relation.
\end{proof}
\end{document}